\documentclass{aa}  

\usepackage{graphicx}
\usepackage{txfonts}
\usepackage{natbib,twoopt}
\usepackage{placeins}
\usepackage{stfloats}
\usepackage[breaklinks=true]{hyperref} 
\bibpunct{(}{)}{;}{a}{}{,}             
\makeatletter
  \newcommandtwoopt{\citeads}[3][][]{\href{http://ui.adsabs.harvard.edu/abs/#3}%
    {\def\hyper@linkstart##1##2{}%
     \let\hyper@linkend\@empty\citealp[#1][#2]{#3}}}
  \newcommandtwoopt{\citepads}[3][][]{\href{http://ui.adsabs.harvard.edu/abs/#3}%
    {\def\hyper@linkstart##1##2{}%
     \let\hyper@linkend\@empty\citep[#1][#2]{#3}}}
  \newcommandtwoopt{\citetads}[3][][]{\href{http://ui.adsabs.harvard.edu/abs/#3}%
    {\def\hyper@linkstart##1##2{}%
     \let\hyper@linkend\@empty\citet[#1][#2]{#3}}}
  \newcommandtwoopt{\citeyearads}[3][][]%
    {\href{http://ui.adsabs.harvard.edu/abs/#3}
    {\def\hyper@linkstart##1##2{}%
     \let\hyper@linkend\@empty\citeyear[#1][#2]{#3}}}
  \newcommandtwoopt{\citedoi}[3][][]{\href{https://doi.org/#3}%
    {\def\hyper@linkstart##1##2{}%
     \let\hyper@linkend\@empty\citealp[#1][#2]{#3}}}
  \newcommandtwoopt{\citepdoi}[3][][]{\href{https://doi.org/#3}%
    {\def\hyper@linkstart##1##2{}%
     \let\hyper@linkend\@empty\citep[#1][#2]{#3}}}
\makeatother

\def\la{\mathrel{\hbox{\rlap{\hbox{\lower4pt\hbox{$\sim$}}}\hbox{$<$}}}}
\def\ga{\mathrel{\hbox{\rlap{\hbox{\lower4pt\hbox{$\sim$}}}\hbox{$>$}}}}

\def\deg      {{\ifmmode^\circ\else$^\circ$\fi} } 
\def\arcmin   {{\ifmmode {'}\else$'$\fi}} 
\def\arcsec   {{\ifmmode{''}\else$''$\fi}} 

\def\Ho       {{$H_{0}$} }

\def\kmsMpc   {{\ $\mathrm{km}\, \mathrm{s}^{-1}\, \mathrm{Mpc}^{-1}$}}

\begin{document} 

   \title{The X-ray properties of the most luminous quasars with strong emission-line outflows}

   \author{Anastasia Shlentsova
          \inst{1,2,3}
          \thanks{\email{ashlentsova@astro.puc.cl}}
          \and          
          Bartolomeo Trefoloni
          \inst{3,4}
          \and
          Matilde Signorini
          \inst{3,5,6}
          \and
          Guido Risaliti
          \inst{2,3}
          \and
          Elisabeta Lusso
          \inst{2,3}
          \and
          Emanuele Nardini
          \inst{3}
          \and       
          Franz E. Bauer
          \inst{7}
          \and       
          Matthew J. Temple
          \inst{8}
          \and       
          Amy L. Rankine
          \inst{9}
          \and       
          Gordon T. Richards
          \inst{10}
          }

   \institute{Instituto de Astrof\'isica, Facultad de F\'isica, Pontificia Universidad Cat\'olica de Chile, Casilla 306, Santiago 22, Chile
         \and
             Dipartimento di Fisica e Astronomia, Universit\`a degli Studi di Firenze, via G. Sansone 1, 50019 Sesto Fiorentino, Firenze, Italy
         \and
             INAF -- Osservatorio Astrofisico di Arcetri, Largo E. Fermi 5, 50125 Firenze, Italy
         \and
             Scuola Normale Superiore, Piazza dei Cavalieri 7, I-56126 Pisa, Italy
         \and
             Dipartimento di Matematica e Fisica, Univerist\`a di Roma 3, Via della Vasca Navale, 84, 00146 Roma RM, Italy
         \and
             European Space Agency (ESA), European Space Research and Technology Centre, Noordwijk, Netherlands
         \and
             Instituto de Alta Investigaci{\'{o}}n, Universidad de Tarapac{\'{a}}, Casilla 7D, Arica, Chile
         \and
             Centre for Extragalactic Astronomy, Department of Physics, Durham University, South Road, Durham DH1 3LE, UK
         \and
             Institute for Astronomy, University of Edinburgh, Royal Observatory, Blackford Hill, Edinburgh EH9 3HJ, UK
         \and
             Department of Physics, Drexel University, 32 S. 32nd Street, Philadelphia, PA 19104, USA
             }

   \date{Received 4 May 2025 / Accepted 19 November 2025}

 
  \abstract
   {Strong outflows from active galactic nuclei are frequently observed in objects with lower coronal X-ray luminosity. This intrinsic X-ray weakness is considered a requirement for the formation of radiatively driven winds.}
   {To obtain an unbiased view on the connection between X-ray emission and the presence of powerful winds in the most luminous quasar phase, we present an X-ray analysis of a sample of extremely luminous, radio-quiet quasars with signatures of strong outflows in their rest-frame ultraviolet (UV) emission spectra.}
   {We study the {\em Chandra} X-ray spectral properties of 10 objects, selected from the Sloan Digital Sky Survey Data Release 16 quasar catalogue based on their UV luminosities and C\,{\sc iv} emission line blueshifts, comparing them to typical optically blue quasars.}
   {Our analysis reveals that seven out of 10 quasars in our sample have photon indices $\Gamma>1.7$. Only two out of 10 objects exhibiting outflows with velocities exceeding 1400~km/s are X-ray `weak', consistent with the fraction of X-ray `weak' objects generally observed in quasar populations. Notably, one of the objects identified as X-ray `weak' is likely an intrinsically X-ray `normal' quasar that is heavily obscured. We observe a tentative indication at a $\sim$2$\sigma$ confidence level that the correlation between the excessively low X-ray flux level and the presence of C\,{\sc iv} emission-line outflows might emerge at wind velocities greater than 3000~km/s.}
   {Our study provides additional evidence that the relationship between X-ray emission and the presence of winds is intricate. Our findings emphasise the need for X-ray observations of a larger sample of UV-selected quasars with confirmed strong emission-line outflows to unravel the nuanced interplay between winds and X-ray emission.}


   \titlerunning{X-ray properties of the most luminous quasars with strong outflows}
   \authorrunning{A. Shlentsova et al.}
   
   \maketitle
%

\section{Introduction}\label{intro}

    Active galactic nuclei (AGNs) are actively accreting supermassive black holes (SMBHs) at the centres of some galaxies. AGN-driven outflows are considered the leading explanation for the co-evolution of SMBHs and their host galaxies (e.g. \citeads{2012ApJ...745L..34Z}; \citeads{2012MNRAS.425..605F}; \citeads{2013ARA&A..51..511K}). Evidence supporting this co-evolution comes from well-established correlations between SMBH mass and various host galaxy properties, including bulge mass, luminosity, and velocity dispersion (e.g. \citeads{1998AJ....115.2285M}; \citeads{2000ApJ...539L...9F}; \citeads{2000ApJ...539L..13G}; \citeads{2002ApJ...574..740T}; \citeads{2004ApJ...604L..89H}). Quasars, the most luminous class of AGNs, are expected to produce outflows powerful enough to rapidly suppress star formation, therefore regulating galaxy evolution (e.g. \citeads{2012A&A...537L...8C}; \citeads{2012ARA&A..50..455F}; \citeads{2015ARA&A..53..115K}; \citeads{2020ApJ...894...28D}, \citeyearads{2024MNRAS.528.4976D}). 
    
    The existence of high-velocity outflows with large column density, arising as a natural consequence of efficient SMBH accretion, is a robust prediction of theoretical models \citepads{2003MNRAS.345..657K} and numerical simulations (\citeads{2013PASJ...65...88T}; \citeads{2023MNRAS.526.3967M}). Observational evidence for these powerful outflows is found in both the ultraviolet (UV) and X-ray spectra. In the UV range, broad absorption lines (BALs) that are blueshifted by several thousand km/s are present in the spectra of $>\!20\%$ of quasars (\citeads{2003AJ....125.1784H}; \citeads{2006ApJS..165....1T}; \citeads{2009ApJ...692..758G}; \citeads{2011MNRAS.410..860A}; \citeads{2023ApJ...952...44B}). Ultra-fast outflows (UFOs) with blueshifts exceeding 10\! \! 000 km/s are observed in the X-ray spectra of $\sim\!30$\,--\,40\% of quasars (\citeads{2010A&A...521A..57T}; \citeads{2013MNRAS.430...60G}). 

    Although AGN feedback via outflows is widely accepted as a mechanism driving the co-evolution of SMBHs and their host galaxies, any conclusive confirmation is still lacking (see \citeads{2017NatAs...1E.165H}, and references therein). Additionally, a definitive understanding of the relationship between the presence of outflows and the steepness of the UV-to-X-ray spectral energy distribution of quasars remains elusive. On the one hand, intrinsic X-ray weakness is generally considered a prerequisite for the presence of winds, as it mitigates over-ionisation and facilitates line driving (\citeads{1995ApJ...451..498M}; \citeads{2000ApJ...543..686P}; \citeads{2004ApJ...616..688P}; \citeads{2014ApJ...789...19H}, \citeyearads{2024MNRAS.527.9236H}; \citeads{2024MNRAS.530.5143D}). On the other hand, evidence suggests that in the Eddington-limited accretion regime, X-ray weakness may also be a consequence of winds (see Figure~13 in \citeads{2023A&A...677A.111T}). For instance, the X-ray-weak sample of \citetads{2019A&A...632A.109N} might be characterised by a higher incidence of excessive UV/optical outflows, as suggested by C\,{\sc iv} emission line profiles with larger velocity shifts and prominent blue wings (\citeads{2021A&A...653A.158L}; see also \citeads{2020A&A...635L...5Z} for additional evidence).
    
    Quasars exhibiting strong absorption signatures from winds are often observed to be X-ray `weak' (\citeads{2002ApJ...567...37G}, \citeyearads{2006ApJ...644..709G}; \citeads{2011MNRAS.413.1013S}). This weakness is typically quantified by comparing the observed X-ray luminosity to that predicted by the X-ray-to-UV luminosity relation observed in `normal' blue quasars (e.g. \citeads{2020A&A...642A.150L}, hereafter L20). Simultaneously, while the features mentioned above provide a direct means of detecting AGN winds, they necessitate a line of sight through the outflowing gas, introducing an inherent bias due to significant modifications in the observed spectra. The faintness of modified spectra makes it unclear whether the observed X-ray weakness is intrinsic or a result of absorption along the line of sight. This ambiguity complicates the inference of how the winds influence the relationship between UV and X-ray luminosities.
    
    The method to circumvent the aforementioned bias and assess the general connection between X-ray emission and the presence of extreme winds in quasars is to utilise emission lines in UV spectra as a means of detecting AGN winds. Several previous studies attempted to infer the influence of the outflow velocities, parametrised by the blueshift of the C\,{\sc iv} broad emission line, on the relationship between UV and X-ray luminosities (e.g. \citeads{2008ApJ...685..773G}; \citeads{2011AJ....142..130K}; \citeads{2018MNRAS.480.5184N}; \citeads{2020A&A...635L...5Z}; \citeads{2020MNRAS.492..719T}; \citeads{2022ApJ...931..154R}; \citeads{2023MNRAS.523..646T}); nevertheless, the findings remain inconclusive. The only way to obtain an unbiased view of the connection between outflows and X-ray properties of quasars is to perform an X-ray analysis of UV-selected sources without any prerequisites on their X-ray characteristics.

    In this paper, we present the X-ray analysis of 10 quasars, selected from the Sloan Digital Sky Survey (SDSS) Data Release 16 (DR16) quasar catalogue \citepads{2020ApJS..250....8L}, which exhibit evidence of strong outflows in their C\,{\sc iv} emission line. We compare these objects with the L20 sample of typical optically selected blue quasars and examine the connection between X-ray emission properties and the presence of powerful winds in the highly luminous quasar phase. The paper is structured as follows. In Section~\ref{sample}, we describe the sample selection, whilst the X-ray data analysis is reported in Section~\ref{analysis}. We present and discuss the results of the X-ray-to-UV luminosity relation fitting in Section~\ref{results}, and conclusions are drawn in Section~\ref{conclusions}.
    
    For simplicity, here we adopt a standard flat $\Lambda$CDM cosmology with $\Omega_{\mathrm{M}}$ = 0.3 and \Ho = 70\kmsMpc.

   \begin{figure}
   \centering
   \includegraphics[width=\hsize]{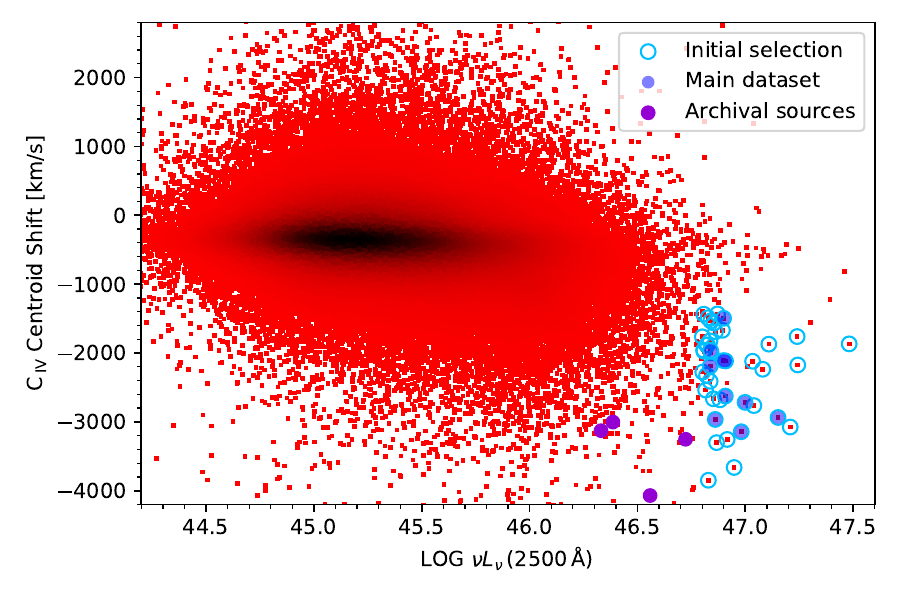}
      \caption{Shift of centroid of C\,{\sc iv} $\lambda$1549~\AA\ line in units of km/s, as function of logarithmic luminosity $\nu L_\nu$ at 2500~\AA, available in \citetads{2022ApJS..263...42W}. The sample consists of all the quasars in the SDSS DR16 sample with redshift in the $z=1.8$\,--\,$2.2$ interval, colour-coded from red to black by the density of the underlying points. Empty cyan circles indicate 45 objects pre-selected for the one-by-one modelling of the C\,{\sc iv} line with multiple components. Filled semi-transparent blue circles show a subset of 10 objects with the confirmed strongly blueshifted component, the main dataset analysed in this paper. Violet circles show four supplementary archival sources.}
         \label{fig:observed}
   \end{figure}


\section{Sample selection}\label{sample}

\subsection{The main dataset}\label{main_data}

   The incidence of blueshifted broad emission lines indicates a prevalence of fast outflows at high luminosities \citepads{2011AJ....141..167R}. This prevalence is further supported by an analysis of the spectral properties of quasars in the SDSS DR16 catalogue, available in \citetads{2022ApJS..263...42W}. Figure~\ref{fig:observed} shows the centroid shift of the C\,{\sc iv} $\lambda$1549~\AA\ emission line as a function of the UV luminosity at 2500~\AA. The centroid shift of the C\,{\sc iv} line is defined as the difference between its measured 50\% flux centroid wavelength (i.e. the wavelength that divides the line into two parts with equal flux) and the rest-frame wavelength. We restricted our analysis to the redshift interval of $z=1.8$\,--\,$2.2$. The lower boundary of $z>1.8$ ensures that the C\,{\sc iv} line is included within the SDSS spectra. The upper boundary of $z<2.2$ is justified by the requirement for the presence of the Mg\,{\sc ii}~$\lambda$2800~\AA\ emission line in the SDSS spectra, which, along with the C\,{\sc iii}]~$\lambda$1909~\AA, provides a reliable redshift measurement independent of the C\,{\sc iv} line. The distribution of the C\,{\sc iv} centroid shifts is roughly symmetric at lower luminosities, whereas significant blueshifts are observed in nearly all sources at the highest luminosities. Therefore, we pre-selected our targets based on their UV luminosity at 2500~\AA\ and the centroid shift of the C\,{\sc iv} emission line.

   We focused on the most luminous, highly accreting quasars with a UV luminosity at 2500~\AA\ of log\,($\nu L_\nu/{\rm erg\,s}^{-1})>46.8$. With a standard bolometric correction from the 2500~\AA\ luminosity of $\sim\!4$--5, this criterion corresponds to a bolometric luminosity exceeding the Eddington luminosity for a $10^9$\,$M_\odot$ black hole. We then selected the sources with a centroid blueshift greater than 1400~km/s (see Figure~\ref{fig:observed}).

\begin{figure}
   \centering
   \includegraphics[width=\hsize]{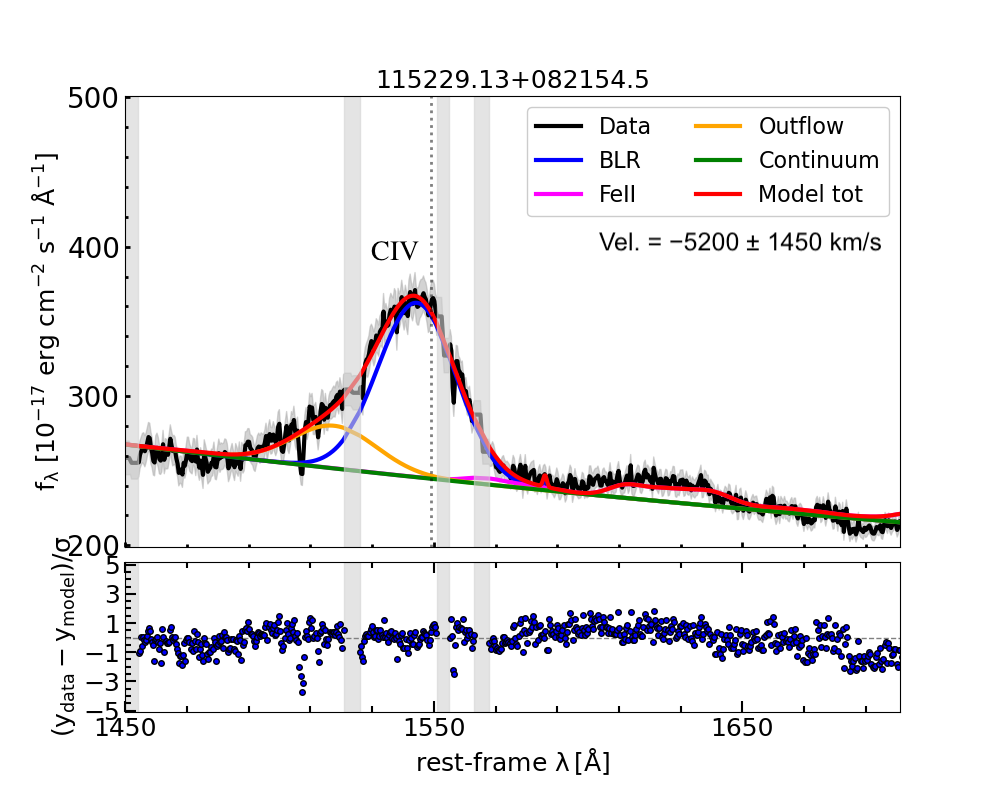}
      \caption{Example of analysis of C\,{\sc iv} spectral region of SDSS spectra. The fitted lines are reported as labels. The components employed in the fit are colour-coded, as shown in the legend. The reported C\,{\sc iv} outflow velocity is measured as the shift of the outflow component with respect to the BLR component. The dashed vertical lines indicate the expected position for the fitted lines according to the redshift reported in the catalogue. The shaded light grey regions are telluric bands, narrow absorption lines, or bad pixels and are therefore excluded from the fit. See Appendix~\ref{optspectrall} for the analysis of other sources.}
         \label{fig:optspectrum}
   \end{figure}

   The SDSS DR16 catalogue does not include information on fitting the C\,{\sc iv} line with multiple Gaussian components, but rather provides the measurements for the whole line profile. Therefore, the selection process outlined above relied on outflow velocities estimated from the centroid shifts of the full C\,{\sc iv} line. Moreover, the distribution of the C\,{\sc iv} blueshifts depends on the combination of the schemes used to determine the quasar redshifts and to quantify the line properties (see Figure~1 in \citeads{2016MNRAS.461..647C}). To confirm that the blueshifts of the 50\% flux centroid are indeed due to the presence of a strongly blueshifted component in the emission line, we conducted a detailed individual analysis of the SDSS spectra of 45 pre-selected sources, modelling the C\,{\sc iv} line with multiple components (see Figure~\ref{fig:optspectrum} and Appendix~\ref{optspectrall} for further details). We also excluded radio-loud quasars from our sample. C\,{\sc iv} outflow velocities, reported in Table~\ref{table:analysis}, were measured in our modelling of the C\,{\sc iv} line as the shift of the outflow component with respect to the broad line region (BLR) component. Since the bulk of the BLR emission for the C\,{\sc iv} line in such bright sources could likely be itself in outflow (see, e.g. \citeads{2018A&A...617A..81V}; and also Figure~\ref{fig:optspectrum}), this approach to calculating the outflow velocity should be regarded as conservative. We additionally calculated outflow velocities non-parametrically as the difference between the wavelength corresponding to the bluest tenth percentile of the line and the wavelength corresponding to the line's peak.

\begin{figure}
   \centering
   \includegraphics[width=\hsize]{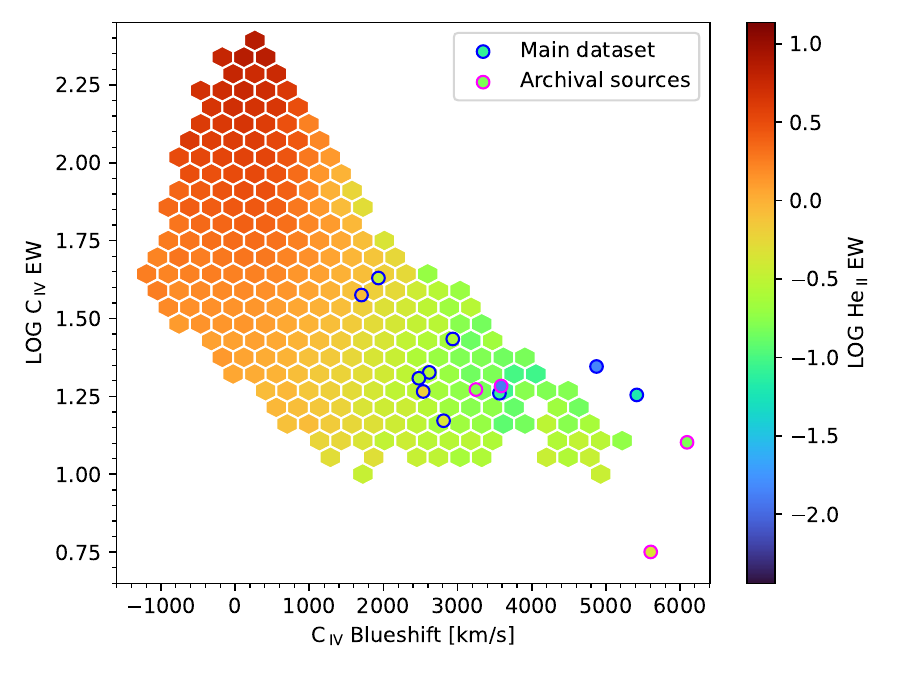}
      \caption{C\,{\sc iv} emission space with hexagons colour-coded by median He\,{\sc ii} EW for 66810 quasars in SDSS DR16 sample with redshift in $z=1.8$\,--\,$2.2$ interval, having reliable C\,{\sc iv} and He\,{\sc ii} measurements from MFICA reconstructions. Only hexagons with five or more quasars are plotted. Circles colour-coded by corresponding He\,{\sc ii} EW show the main dataset of 10 objects and four supplementary archival sources.}
         \label{fig:observed_MFICA}
   \end{figure}

   We further confirmed the presence of outflows using C\,{\sc iv} emission line blueshifts calculated from spectral reconstructions based on a mean-field independent component analysis (MFICA) scheme (\citeads{2020MNRAS.492.4553R}; \citeads{2023MNRAS.523..646T}). In this approach, the C\,{\sc iv} blueshift is defined as the Doppler shift of the wavelength bisecting the continuum-subtracted line flux (i.e. the rest-frame wavelength of the observed line centroid) with respect to the mean rest-frame wavelength of the C\,{\sc iv} doublet, obtained using the improved redshift estimation. Improved redshifts were obtained from P.~Hewett (private communication and Hewett et al. 2026). The scheme used 27 high signal-to-noise ratio templates in a fashion similar to that described in Section~2.1 of \citetads{2023MNRAS.524.5497S}. A key difference, however, is that the templates and spectra were cross-correlated (see Equation~1 of \citeads{2010MNRAS.405.2302H}) using a wavelength range restricted to rest-frame 1700\,--3000\,\AA. Constraining the match to include only the C\,{\sc iii}]~$\lambda$1909~\AA\ emission complex and Mg\,{\sc ii}~$\lambda$2800~\AA\ emission avoids biases that arise from the presence of C\,{\sc iv} absorbers in the blue wing of the C\,{\sc iv} emission line. All objects selected in the aforementioned process have C\,{\sc iv} blueshifts measured from MFICA reconstructions greater than 1700~km/s. Figure~\ref{fig:observed_MFICA} shows MFICA reconstructed C\,{\sc iv} blueshifts and equivalent widths (EWs), and He\,{\sc ii} EWs of the selected objects in relation to those of the $z=1.8$\,--\,$2.2$ quasars in the SDSS DR16 sample.
   
   The main dataset presented in this paper consists of 10 sources. Table~\ref{table:newobs} provides the corresponding observing parameters of the X-ray data. Source SDSS~J073502.30+265911.5 was observed by the {\em Chandra} X-ray Observatory in October 2015 (cycle 16, proposal ID: 16700345, PI: E. Piconcelli). The Advanced CCD Imaging Spectrometer (ACIS) instrument operated in VFAINT TE mode, with an on-source time of 24.75 ks. This object is part of the WISSH quasars sample \citepads{2017A&A...608A..51M}, but we performed a complete reprocessing and data analysis to ensure the homogeneity of measurements across the main dataset. The {\em Chandra} observations of the remaining nine quasars in our sample are proprietary and previously unpublished. Observations began in December 2023 and were completed in August 2025 (cycle 25, proposal ID: 25700372, PI: G. Risaliti). For each target, the ACIS instrument operated in FAINT TE mode, with on-source times ranging from 10.93 to 18.83 ks. The observation data files were reprocessed using the Chandra Interactive Analysis of Observations (CIAO) software package \citepads{2013ascl.soft11006C} v4.16.0 and the Chandra calibration database (CalDB) v4.11.2, applying the point-source aperture correction to the unweighted auxiliary response files. 

\begin{table*}
{\tiny
\caption{Observing parameters of X-ray data.}             
\label{table:newobs}      
\centering          
\begin{tabular}{l c c c c c}
\hline\hline       
SDSS name (J-) & Instrument & Observation ID & Observation Start Date & Exposure Time (ks) & Count Rate (0.3--8 keV)\\
(1) & (2) & (3) & (4) & (5) & (6) \\
\hline                    
 \multicolumn{6}{c}{Main Dataset} \\
\hline                 
    073502.30+265911.5 & {\em Chandra} & 17077 & 2015 Oct 02 & 24.75 & $7.7 \pm 0.6$\\
    081812.68+181913.6 & {\em Chandra} & 29189 & 2024 Jan 10 & 10.93 & $3.0 \pm 0.5$\\
    082450.79+154318.4 & {\em Chandra} & 28286 & 2024 Jan 08 & 16.85 & $9.1 \pm 0.7$\\
    083046.17+152329.7 & {\em Chandra} & 28292 & 2024 Jan 04 & 18.83 & $11.0 \pm 0.8$\\
    090924.01+000211.0 & {\em Chandra} & 28288 & 2024 Apr 25 & 16.85 & $13.1 \pm 0.9$\\
    093514.71+033545.7 & {\em Chandra} & 28289 & 2024 May 05 & 17.84 & $3.0 \pm 0.4$\\
    103005.10+132531.1 & {\em Chandra} & 28287 & 2024 Feb 29 & 16.85 & $4.0 \pm 0.5$\\
    111800.50+195853.4 & {\em Chandra} & 28285 & 2024 Jun 26 & 16.66 & $1.2 \pm 0.3$\\
    115229.13+082154.5 & {\em Chandra} & 28290 & 2023 Dec 20 & 17.70 & $2.6 \pm 0.4$\\
    133610.96+184529.9 & {\em Chandra} & 28284 & 2025 Aug 18 & 15.86 & $0.2 \pm 0.1$\\
\hline                  
 \multicolumn{6}{c}{Supplementary Archival Data} \\
\hline                  
   082508.75+115536.3\, \tablefootmark{a} & {\em Chandra} & 14951 & 2013 Jun 10 & 5.10 & --\\
   092156.38+285237.7\, \tablefootmark{b} & {\em XMM-Newton} & 0822530301 & 2018 May 15 & 32.7 & $19.0 \pm 1.0$\\
   104350.35+140703.0\, \tablefootmark{b} & {\em XMM-Newton} & 0904720601 & 2022 Jun 10 & 11.4 & $1.9 \pm 0.4$\\
   122048.52+044047.6\, \tablefootmark{c} & {\em Chandra} & 18112 & 2016 Jan 20 & 3.37 & --\\
\hline                  
\end{tabular}
\tablefoot{Columns: (1) source name in the SDSS DR16 catalogue; (2)\,--\,(4) instrument, observation ID, and observation start date; (5) background-flare cleaned effective exposure time, in ks; (6) net count rate in the 0.3--8 keV band, in $10^{-3}$ counts s$^{-1}$. Archival data obtained from: 
\tablefoottext{a}{\citetads{2015ApJ...805..122L}};
\tablefoottext{b}{\citetads{2020A&A...641A.137T}};
\tablefoottext{c}{\citetads{2018MNRAS.480.5184N}}.
}
}
\end{table*}

\begin{table*}
{\tiny
\caption{Optical properties and results of C\,{\sc iv} region and X-ray spectral analysis.}             
\label{table:analysis}      
\centering          
\begin{tabular}{l c c c c c c c c c c}
\hline\hline       
SDSS name (J-) & $z$ & $M_i$ & $N_{\mathrm{H, Gal}}$ & $\mathrm{log}\ L_{2500\, \mathrm{Å}}$ & C\,{\sc iv} Velocity & $\Gamma$ & $\mathrm{log}\ F_{1-10\, \mathrm{keV}}$ & $\mathrm{log}\ L_{2\, \mathrm{keV}}$ & $\alpha_{\mathrm{OX}}$ & $\Delta\alpha_{\mathrm{OX}}$\\
(1) & (2) & (3) & (4) & (5) & (6) & (7) & (8) & (9) & (10) & (11) \\
\hline                    
 \multicolumn{11}{c}{Main Dataset} \\
\hline                    
    073502.30+265911.5 & 1.999 & $-29.71$ & 5.38 & 32.07 & $-4900 \pm 1000$ & $1.6 \pm 0.2$ & $-13.1 \pm 0.1$ & $27.1 \pm 0.1$ & $-1.91 \pm 0.04$ & $-0.21$\\
    081812.68+181913.6 & 2.010 & $-29.11$ & 3.00 & 31.82 & $-6170 \pm 460$ & $2.2 \pm 0.8$ & $-13.3 \pm 0.2$ & $27.2 \pm 0.4$ & $-1.76 \pm 0.16$ & $-0.08$\\
    082450.79+154318.4 & 1.886 & $-28.94$ & 2.83 & 31.83 & $-4040 \pm 540$ & $2.6 \pm 0.3$ & $-12.9 \pm 0.1$ & $27.7 \pm 0.1$ & $-1.57 \pm 0.05$ & 0.11\\
    083046.17+152329.7 & 1.955 & $-28.91$ & 3.27 & 31.78 & $-3140 \pm 160$ & $2.1 \pm 0.2$ & $-12.8 \pm 0.1$ & $27.6 \pm 0.1$ & $-1.59 \pm 0.05$ & 0.08\\
    090924.01+000211.0 & 1.878 & $-29.17$ & 2.56 & 31.82 & $-3980 \pm 690$ & $1.9 \pm 0.2$ & $-12.7 \pm 0.1$ & $27.6 \pm 0.1$ & $-1.63 \pm 0.05$ & 0.05\\
    093514.71+033545.7 & 1.832 & $-28.88$ & 2.37 & 31.76 & $-3480 \pm 630$ & $2.5 \pm 0.5$ & $-13.4 \pm 0.1$ & $27.2 \pm 0.2$ & $-1.74 \pm 0.10$ & $-0.07$\\
    103005.10+132531.1 & 1.887 & $-28.95$ & 3.28 & 31.83 & $-6780 \pm 240$ & $1.9 \pm 0.4$ & $-13.2 \pm 0.1$ & $27.1 \pm 0.2$ & $-1.80 \pm 0.09$ & $-0.13$\\
    111800.50+195853.4 & 1.962 & $-29.28$ & 1.15 & 31.90 & $-4410 \pm 1070$ & $1.2 \pm 0.8$ & $-13.6 \pm 0.2$ & $26.2 \pm 0.5$ & $-2.17 \pm 0.19$ & $-0.48$\\
    115229.13+082154.5 & 1.868 & $-28.69$ & 1.46 & 31.76 & $-5200 \pm 1450$ & $2.4 \pm 0.6$ & $-13.5 \pm 0.1$ & $27.1 \pm 0.3$ & $-1.78 \pm 0.11$ & $-0.12$\\
    133610.96+184529.9 & 1.948 & $-29.09$ & 1.55 & 31.92 & $-3680 \pm 280$ & ... & $<-13.8$ & $<26.6$ & $< -2.05$ & $<-0.37$\\
\hline                  
 \multicolumn{11}{c}{Supplementary Archival Data} \\
\hline                  
   082508.75+115536.3 & 1.996 & $-28.71$ & 4.09 & 31.64 & $-4560 \pm 1020$ & ...\, \tablefootmark{a} & -- & $< 26.4$\, \tablefootmark{a} & $<-2.02$ & $<-0.37$\\
   092156.38+285237.7 & 1.730 & $-27.96$ & 1.65 & 31.48 & $-2020 \pm 970$ & $2.5 \pm 0.1$ & $-13.1\pm0.1$ & $27.6\pm0.1$ & $-1.48\pm0.05$ & 0.15\\
   104350.35+140703.0 & 2.321 & $-27.53$ & 2.72 & 31.25 & $-6710 \pm 460$ & $2.0 \pm 0.4$ & $-14.0 \pm 0.2$ & 26.5$\pm0.3$ & $-1.81\pm0.12$ & $-0.21$\\
   122048.52+044047.6 & 1.736 & $-27.76$ & 1.59 & 31.31 & $-6420 \pm 3160$ & $> 1.4$\, \tablefootmark{b} & -- & 27.1\, \tablefootmark{b} & $-1.63$ & $-0.02$\\
\hline                  
\end{tabular}
\tablefoot{Columns: (1) source name in the SDSS DR16 catalogue; (2) redshift from \citetads{2022ApJS..263...42W}; (3) absolute \textit{i}-band magnitude from \citetads{2020ApJS..250....8L}; (4) Galactic neutral hydrogen column density from HEASARC (\citeads{1990ARA&A..28..215D}; \citeads{2005A&A...440..775K}; \citeads{2016A&A...594A.116H}), in $10^{20}$ cm$^{-2}$; (5) monochromatic luminosity at rest-frame 2500~\AA\ from \citetads{2022ApJS..263...42W}, in erg~s$^{-1}$; (6) outflow velocity, measured in our modelling of the C\,{\sc iv} line as the shift of the outflow component with respect to the BLR component, in km/s; (7) photon index of the X-ray continuum in the Galactic absorption-corrected resulting model, an entry of ``...'' indicates that it cannot be constrained; (8) Galactic absorption-corrected observed-frame 1–10~keV flux as inferred from the resulting model, in erg~cm$^{-2}$~s$^{-1}$; (9) intrinsic monochromatic luminosity at rest-frame 2~keV, in erg~s$^{-1}$; (10) measured $\alpha_{\mathrm{OX}}$ parameter, defined as $\alpha_{\mathrm{OX}} = -0.3838\, \mathrm{log}(L_{2500\, \mathrm{Å}}/L_{2\, \mathrm{keV}})$; (11) difference between the measured $\alpha_{\mathrm{OX}}$ parameter and the expected $\alpha_{\mathrm{OX}}$ parameter from the best-fit $\alpha_{\mathrm{OX}}$ -- $L_{\mathrm{UV}}$ relation for the L20 sample, defined as $\alpha_{\mathrm{OX}} = (-0.134 \pm 0.003)\, \mathrm{log}\, L_{2500\, \mathrm{Å}} + (2.586 \pm 0.093)$. Upper limits indicate non-detection. Values of $\Gamma$ and $F_{2\, \mathrm{keV}}$ used to calculate $L_{2\, \mathrm{keV}}$ for the archival sources obtained from
\tablefoottext{a}{\citetads{2015ApJ...805..122L}};
\tablefoottext{b}{\citetads{2018MNRAS.480.5184N}}.
}
}
\end{table*}

\subsection{Supplementary archival data}\label{sup_data}

   We find no significant evidence for a correlation between the excessively low X-ray flux level of quasars and the presence of outflows in our main dataset (see Section~\ref{relation} for details). Therefore, we also considered a control sample to check whether any such correlation emerges at a higher velocity threshold.
   
   We searched through the {\em Chandra} and {\em XMM-Newton} archives for observed highly luminous quasars that exhibit a C\,{\sc iv} centroid blueshift, calculated from the spectral properties reported by \citetads{2022ApJS..263...42W}, greater than 3000~km/s. We expanded the redshift interval to $z=1.7$\,--\,$2.4$, relaxed our UV luminosity criterion to log\,($\nu L_\nu/{\rm erg\,s}^{-1})>46.3$ at 2500~\AA, and identified four additional sources with available X-ray observations (see Table~\ref{table:newobs} for the observing parameters). SDSS~J082508.75+115536.3 and SDSS~J122048.52+044047.6 were observed with {\em Chandra} (\citeads{2015ApJ...805..122L}; \citeads{2018MNRAS.480.5184N}). Their rest-frame monochromatic luminosities at 2~keV were calculated based on the reported rest-frame 2-keV flux densities (see Table~\ref{table:analysis}). We acknowledge that SDSS~J082508.75+115536.3 was targeted as a likely X-ray `weak' object and SDSS~J122048.52+044047.6 as a potentially X-ray `weak' one, which possibly introduces a~bias into our control sample. SDSS~J092156.38+285237.7 and SDSS~J104350.35+140703.0 were located in the 4XMM-DR14 Serendipitous Source Catalogue \citepads{2020A&A...641A.137T}. In order to obtain a homogeneous analysis, we reduced these two {\em XMM-Newton} sources following the standard procedure outlined in the European Space Agency's {\em XMM-Newton} web pages, and we performed a spectral analysis analogous to the one described for {\em Chandra} sources (see Section~\ref{analysis} for details).
   
   MFICA reconstructed C\,{\sc iv} blueshifts of these supplementary archival objects are greater than 3200~km/s. We further confirmed the presence of a strongly blueshifted component in the C\,{\sc iv} emission line and measured outflow velocities by modelling the line with multiple components, similar to the sources in our main dataset.


\section{X-ray data analysis}\label{analysis}

   \begin{figure}
   \centering
   \includegraphics[width=\hsize]{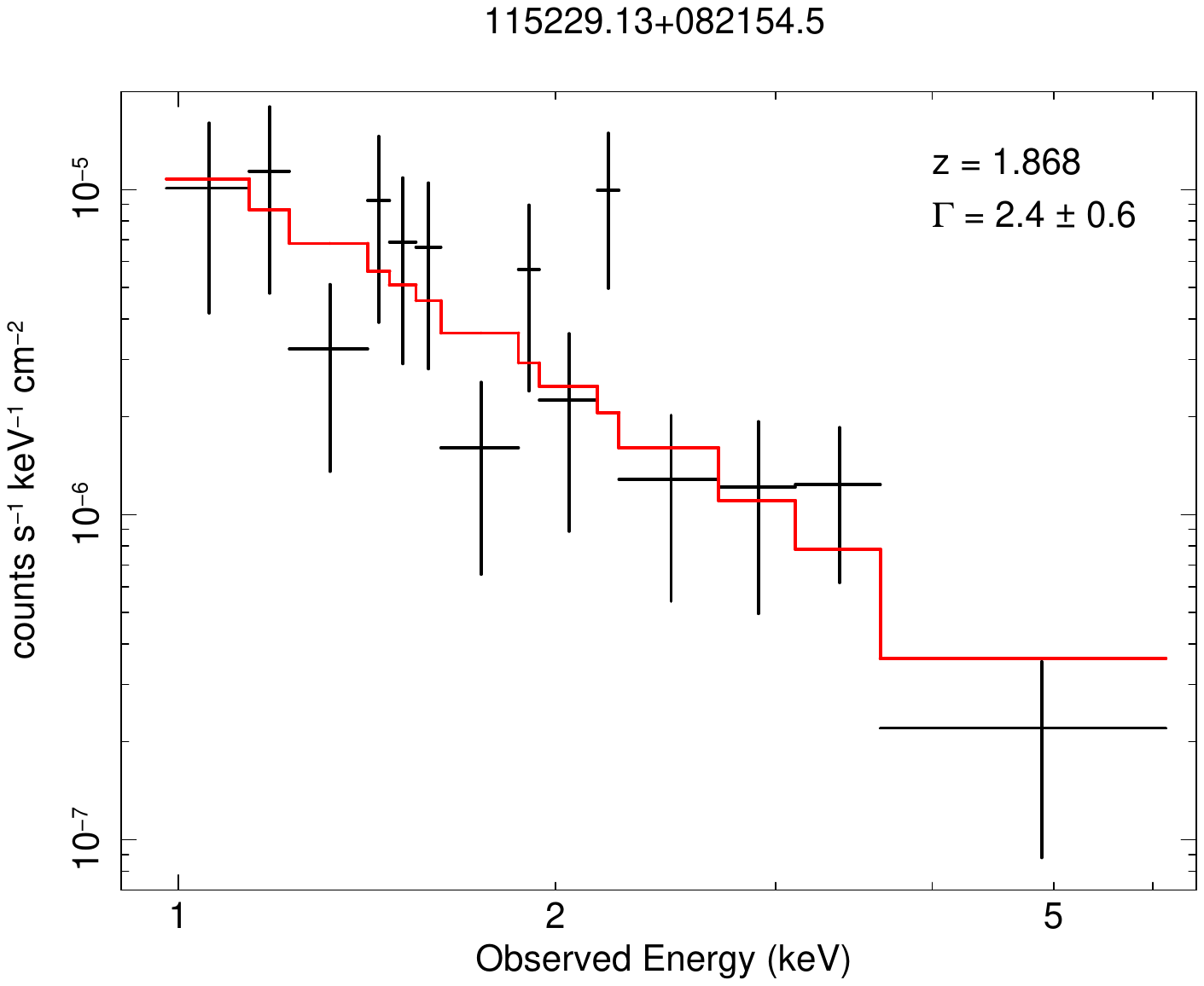}
      \caption{Example of analysis of {\em Chandra} X-ray spectra. To enhance visual clarity, the spectrum in the plot is binned to ensure at least 1.5 counts per energy channel. Black crosses represent the observational data with corresponding errors, red line represents the best-fit model. See Appendix~\ref{Xspectrall} for the analysis of other sources.
              }
         \label{fig:Xspectrum}
   \end{figure}

   The spectral analysis of the main dataset was conducted using the general X-ray spectral-fitting program \verb|XSPEC| \citepads{1999ascl.soft10005A} v12.14.0. We re-binned the data to ensure that there was at least one count per energy channel and employed the C-statistic (\citeads{1979ApJ...228..939C}; \citeads{2017A&A...605A..51K}), which is more appropriate for the low-count Poissonian regime. The reported uncertainties correspond to a change in the fit statistics of $\Delta C = 2.706$ (equivalent to the 90\% confidence level in the Gaussian approximation).
   
   Despite the limited quality of the obtained spectra, we achieved a robust estimation of the X-ray flux for all sources in our sample, except for one with a non-detection (see Figure~\ref{fig:Xspectrum} and Appendix~\ref{Xspectrall}). The spectra were fitted over the 0.3--8~keV energy range, as virtually no source counts were detected outside this range. The initially adopted spectral model consisted of a power-law continuum modified by Galactic absorption from \citetads{2016A&A...594A.116H}, as well as absorption in the vicinity of the sources. In \verb|XSPEC| terminology, this is expressed as \verb|phabs| $\times$ \verb|zphabs| $\times$ \verb|powerlw|. We found that the non-Galactic H\,{\sc i} column density is $\mathrm{log}\ N_{\mathrm{H}} < 21.5$, with $\Delta C = 3.84$ (equivalent to the 95\% confidence level in the Gaussian approximation), for all sources in our main dataset, except for one with a non-detection. Therefore, for further analysis, we fixed the non-Galactic absorption parameter to zero. Consequently, our resulting spectral model included only two free parameters: the photon index of the power law and the flux, assessed through \verb|cflux| in \verb|XSPEC| (which has been omitted from the model definition above for simplicity).

   For each source in the main dataset with a detection, we determined the observed-frame 1–10~keV flux and estimated the monochromatic rest-frame 2-keV luminosity. The inferred power-law photon indices, measured fluxes, and estimated luminosities are provided in Table~\ref{table:analysis}. In the case of SDSS~J111800.50+195853.4, we additionally estimated the rest-frame 2-keV luminosity assuming the source to be an X-ray steep but heavily obscured quasar, by allowing the non-Galactic absorption parameter to vary while fixing the photon index at $\Gamma=2.0$, the average photon index for a blue quasar (e.g. \citeads{2009A&A...495..421B}; \citeads{2009ApJS..183...17Y}). This analysis resulted in the non-Galactic H\,{\sc i} column density of $\mathrm{log}\ N_{\mathrm{H}} = 23.2^{+0.4}_{-1.4}$ and monochromatic rest-frame 2-keV luminosity of $\mathrm{log}\ L_{2\, \mathrm{keV}} = 27.8 \pm 0.2$. In the case of SDSS~J133610.96+184529.9, the only source for which the count rate in the 0.3--8~keV energy range corresponds to a non-detection, we fixed the model by setting the non-Galactic absorption parameter to zero and the photon index to $\Gamma=2.0$. With the resulting model having only the flux as a free parameter, we estimated upper limits of the observed-frame 1–10~keV flux and the monochromatic rest-frame 2-keV luminosity at a 90\% confidence level. The measured values are consistent with those obtained using confidence limits from \citetads{1991ApJ...374..344K}.
   

\section{Results}\label{results}
\subsection{Photon index distribution}\label{gamma}

   It is essential to consider the possibility of intrinsic X-ray absorption when investigating the X-ray properties of quasars. In order to disentangle intrinsic X-ray weakness from the presence of absorption, particularly in cases where the fit returns low non-Galactic H\,{\sc i} column densities, the continuum photometric photon index, $\Gamma$, can be utilised \citepads{2022ApJ...931..154R}. Objects with flatter X-ray spectra than the average (i.e. low photon index) are likely affected by absorption, while those observed as X-ray `weak' with relatively steep X-ray spectra are likely intrinsically X-ray weak (see Figures~1 and 2 in \citeads{2019A&A...630A..94G}). Therefore, we compare the continuum photon indices of objects in our main dataset with those in the L20 parent sample.
   
   In the L20 parent sample, prior to applying the photon index selection criterion, the distribution of photon indices has a mean value of 1.94 and a standard deviation of 0.45. Our main dataset has a distribution of photon indices with a mean value of 2.04 and a standard deviation of 0.42. These properties of the distribution indicate that, on average, our sources reside among quasars with steep X-ray spectra, suggesting very low intrinsic X-ray absorption. Notably, we find that seven out of 10 objects in our sample have X-ray spectra with photon indices $\Gamma>1.7$ (see Table~\ref{table:analysis}). Considering that quasars in our sample exhibit a strongly blueshifted C\,{\sc iv} emission line with moderate to low EW (see Figure~\ref{fig:observed_MFICA}), this result is consistent with previous studies indicating that objects with high C\,{\sc iv} blueshifts and low C\,{\sc iv} EWs tend to have steep X-ray spectra (e.g. \citeads{2005AJ....129..567G}; \citeads{2011AJ....142..130K}; \citeads{2022ApJ...931..154R}).   
   
\subsection{$L_{\mathrm{X}}$ -- $L_{\mathrm{UV}}$ relation}\label{relation}

   \begin{figure}
   \centering
   \includegraphics[width=\hsize]{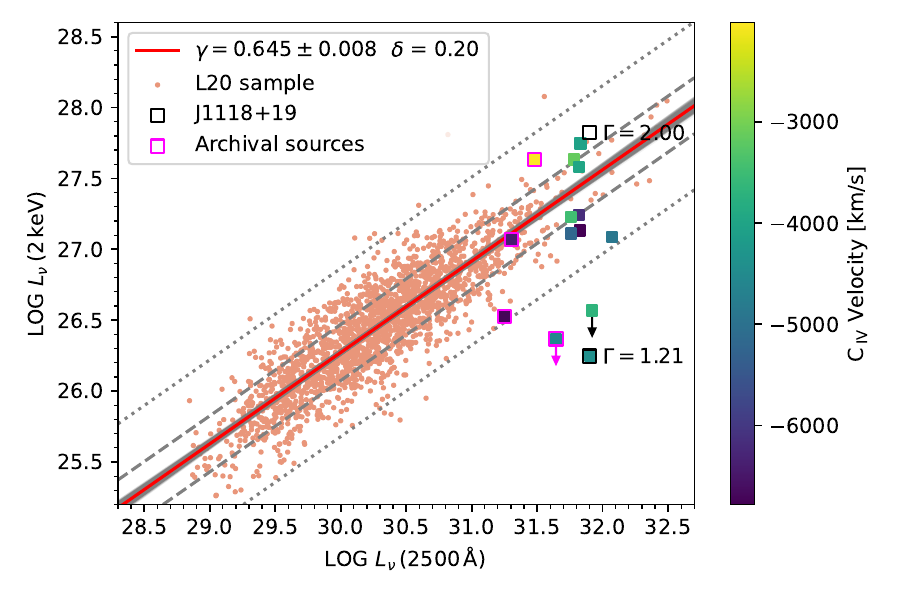}
      \caption{Rest-frame monochromatic luminosities $L_{\mathrm{X}}$ against $L_{\mathrm{UV}}$ for 10 quasars in our main dataset and four additional quasars from {\em Chandra} and {\em XMM-Newton} archives, colour-coded by C\,{\sc iv} outflow velocity in units of km/s, measured in our modelling of C\,{\sc iv} line as the shift of the outflow component with respect to the BLR component. Colour-coded and empty black squares show modelling of J111800.50+195853.4 with a free and fixed photon index, respectively. Downward arrows indicate non-detection, and the corresponding squares represent upper limits. Light red dots represent the sample of about 2000 quasars from \citetads{2020A&A...642A.150L}, with the relative regression line in red, for which the slope $\gamma$, with its error, and the dispersion $\delta$ are specified. The dashed and dotted lines trace the $1\sigma$ and $3\sigma$ dispersion, respectively.}
         \label{fig:relation}
   \end{figure}

   The L20 sample is suitable for cosmological applications that utilise a quasar Hubble diagram, where luminosity distances are derived from the X-ray-to-UV luminosity relation (e.g. \citeads{2015ApJ...815...33R}; \citeyearads{2019NatAs...3..272R}; \citeads{2019A&A...628L...4L}). Due to the selection criteria in L20, aimed at isolating the quasars with reliable measurements of intrinsic UV and X-ray emission, the dispersion in the $L_{\mathrm{X}}$ -- $L_{\mathrm{UV}}$ relation obtained from that sample is as low as 0.2~dex. Therefore, the L20 sample can be regarded as consisting of X-ray `normal' optically blue quasars. Consequently, we evaluated the X-ray intensity of the objects in our sample by comparing their luminosities to those expected from the $L_{\mathrm{X}}$ -- $L_{\mathrm{UV}}$ relation established for the L20 sample.
   
   We determined the slope $\gamma$, intercept, and dispersion $\delta$ of the X-ray-to-UV luminosity relation for the L20 sample using the Python package \verb|EMCEE| \citepads{2013ascl.soft03002F}, a pure-Python implementation of Goodman \& Weare’s affine invariant Markov chain Monte Carlo (MCMC) ensemble sampler. The regression fit was conducted with the normalisation of X-ray and UV luminosities set to the respective median values of the analysed sample.

   Figure~\ref{fig:relation} demonstrates that eight out of 10 quasars in our main dataset are X-ray `normal' within $\Delta C = 3.84$ and the $3\sigma$ dispersion of the $L_{\mathrm{X}}$ -- $L_{\mathrm{UV}}$ relation. The only two X-ray `weak' quasars are J111800.50+195853.4 and J133610.96+184529.9. J111800.50+195853.4 is an outlier of the X-ray-to-UV luminosity relation at $6\sigma$. J133610.96+184529.9 has an upper limit below the X-ray-to-UV luminosity relation at $5\sigma$. For seven objects in our main dataset with $\Gamma>1.7$, we additionally examined the X-ray-to-UV luminosity relation in narrow redshift bins. We confirmed that these objects are X-ray `normal' within $\Delta C = 2.706$ and the $2\sigma$ dispersion of the relation in their respective redshift bins (see Appendix~\ref{relation_in_bins} for further details). Therefore, our main dataset of the most luminous quasars contains two X-ray `weak' objects out of 10, which aligns with the average expectation for quasar populations. Cautious about the small size of our sample, we estimated the statistical uncertainty for the fraction of X-ray `weak' objects using the Clopper-Pearson method \citepdoi{10.1093/biomet/26.4.404}, which conservatively guarantees that the interval coverage is always equal to or above the confidence level (e.g. \citedoi{10.1214/ss/1009213286}). The $68\%$ confidence interval for the true fraction is $10.6$\,--\,32.0$\%$. Furthermore, if we assume J111800.50+195853.4 to be a heavily obscured quasar with an intrinsic continuum photon index of $\Gamma=2.0$, its estimated monochromatic rest-frame 2-keV luminosity becomes consistent with that of X-ray `normal' quasars within the $2\sigma$ dispersion of the $L_{\mathrm{X}}$ -- $L_{\mathrm{UV}}$ relation, bringing the mean fraction of intrinsically X-ray `weak' objects down to one out of 10 and the $68\%$ confidence interval for the true fraction to $1.6$\,--\,26.3$\%$.

   Considering the high uncertainty in the C\,{\sc iv} outflow velocities measured in our analysis of the SDSS spectra for some sources in our sample (see Table~\ref{table:analysis}), we repeated the examination of a possible correlation between the X-ray flux level of quasars and the presence of outflows using different methods to measure the outflow velocity. We used outflow velocities that we measured non-parametrically, shifts of the centroid of the C\,{\sc iv} line, reported by \citetads{2022ApJS..263...42W}, and those measured from MFICA reconstructions. Despite the discrepancy between the values of the outflow velocity measured using different methods for each particular source, our results remain consistent, showing no significant evidence of a correlation between the excessively low X-ray flux level and the presence of strong winds in the most luminous quasar phase. At the same time, the position of our sample within the C\,{\sc iv} emission space (see Figure~\ref{fig:observed_MFICA}) corresponds to the regime dominated by an efficient radiation line-driven disc wind, expected to occur in cases of high black hole mass and high mass-normalised accretion rates \citepads{2023MNRAS.523..646T}, which aligns with the luminosity characteristics of our sample. Relatively low He\,{\sc ii} EWs of quasars in our sample suggest a weak extreme UV ionising emission compared to the UV continuum \citepads{2021MNRAS.504.5556T}, which is required to avoid over-ionising the gas responsible for line driving. Therefore, He\,{\sc ii} EWs support the possibility for our objects to efficiently launch a radiation-driven wind without the necessity for excessive steepness of the UV-to-X-ray spectral energy distribution \citepads{2023MNRAS.523..646T}. Our findings align with previous studies (e.g. \citeads{2020MNRAS.492..719T}; \citeads{2022ApJ...931..154R}), providing supplementary evidence of the complexity of the relationship between X-ray emission and the presence of winds in luminous quasars. 
   
   It is worth noting that this absence of correlation between the excessive steepness of the UV-to-X-ray spectral energy distribution and the presence of strong winds could be explained by the non-simultaneity of the X-ray and UV observations used in our analysis. Our objects could have been X-ray `weak' during the outflow phase, which was concealed by variations in X-ray and UV properties between the corresponding observations due to intrinsic variability in quasars. Alternatively, the reason may lie in the difference between the ionising continuum seen by the BLR and in our observations (see, e.g. Section~2.3 in \citeads{2004ApJ...611..125L}), or the fact that the C\,{\sc iv} blueshifts have some orientation dependence rather than solely reflecting outflow velocities (see, e.g. \citeads{2021ApJ...914L..14R}). The relative locations of the accretion disk, the BLR, and the X-ray corona with respect to each other and our line of sight could influence the observed C\,{\sc iv} emission line profiles, at the same time contributing to the X-ray weakness at higher inclinations.

   Finally, we evaluated the probability that the correlation between the excessively low X-ray intensity in the most luminous quasar phase and the presence of strong outflows might emerge at a higher velocity threshold. We supplemented our main dataset with archival data and divided the sample into two subsets. The first subset includes five quasars exhibiting blueshift of the centroid of the C\,{\sc iv} line, reported by \citetads{2022ApJS..263...42W}, greater than 3000~km/s, of which two objects we identified as X-ray `weak'. The second subset includes the remaining nine quasars, of which only one was identified as X-ray `weak'. The probability that the two subsets have the same underlying fraction of X-ray `weak' sources is $P_{\mathrm{null}}=0.012$. This null hypothesis probability corresponds to a 2.5$\sigma$ confidence level for the presence of a correlation between the excessively low X-ray flux level and outflows faster than 3000~km/s. If we assume J111800.50+195853.4 to be a heavily obscured X-ray `normal' quasar, the null hypothesis probability becomes $P_{\mathrm{null}}=0.062$, corresponding to a 1.9$\sigma$ confidence level for the presence of a correlation. Therefore, we do not claim any correlation between the X-ray properties and the presence of strong outflows in the most luminous quasar phase based on our small sample, but note that there is a hint at a $\sim$2$\sigma$ level that the correlation might emerge for winds faster than 3000~km/s. A larger sample is required to understand whether the correlation between the two phenomena is due to a statistical fluctuation or a physical effect.
   

\section{Conclusions}\label{conclusions}

    We have presented an X-ray analysis of a sample of 10 extremely luminous, radio-quiet, non-BAL quasars with evidence of strong outflows in their rest-frame UV emission spectra. This sample was selected from the SDSS DR16 quasar catalogue based on optical/UV characteristics, without any prerequisites regarding X-ray properties. The presence of strong outflows, parametrised by the blueshift of the C\,{\sc iv} broad emission line, was confirmed in a one-by-one analysis of the SDSS spectra of the sources. We analysed the X-ray properties of the sample and compared them with those of a sample of typical optically selected blue quasars. Our main findings can be summarised as follows:
    \begin{enumerate}
      \item The photon indices for quasars in our main dataset show that our sources mostly exhibit X-ray steep spectra. We find that seven out of 10 objects have photon indices $\Gamma>1.7$, and are therefore unlikely to be absorbed.
      \item Only two out of 10 highly luminous quasars with C\,{\sc iv} emission-line outflows exceeding 1400~km/s are X-ray `weak', compared to predictions from the X-ray-to-UV luminosity relation. This result is consistent with the average fraction observed in quasar populations overall. One of the objects identified as X-ray `weak' is possibly an intrinsically X-ray `normal', yet heavily obscured, quasar.
      \item We observe a tentative indication of the correlation between the excessively low X-ray intensity and the presence of C\,{\sc iv} emission-line outflows that exceed 3000~km/s at a $\sim$2$\sigma$ confidence level. However, a larger sample is required to determine whether this finding results from a statistical fluctuation or reflects a physical effect.
   \end{enumerate}
   The steepness of X-ray spectra in our sample of extremely luminous quasars with C\,{\sc iv} emission-line outflows exceeding 1400~km/s, along with the fact that most of our objects follow the $L_{\mathrm{X}}$ -- $L_{\mathrm{UV}}$ relation of typical optically blue quasars, is consistent with previous studies. Our results contribute to the body of evidence indicating that the relationship between X-ray emission and the presence of winds in luminous quasars is complicated and perplexing.
   
   Given the probabilistic nature of our results, extending X-ray observations to larger samples of UV-selected quasars with confirmed high-velocity outflows will be important. These observations will help disentangle the different processes affecting the C\,{\sc iv} blueshifts and clarify the complex interplay between outflows and X-ray emission in quasars.


\begin{acknowledgements}
      This research has utilised data obtained from the 4XMM XMM-Newton serendipitous stacked source catalogue 4XMM-DR14, compiled by the institutes of the XMM-Newton Survey Science Centre selected by ESA. The authors are grateful to the anonymous referee for useful and constructive comments. We thank the Munich Institute for Astro-, Particle, and BioPhysics (MIAPbP), which is funded by the Deutsche Forschungsgemeinschaft (DFG, German Research Foundation) under Germany's Excellence Strategy -- EXC-2094 -- 390783311, for providing space for fruitful discussions on this work. AS is supported by the national doctoral scholarship from the Agencia Nacional de Investigaci\'on y Desarrollo (ANID), folio de postulaci\'on 21221788. AS personally acknowledges Olesya and Kirill Kuchay for their unwavering support. BT acknowledges support by the European Union’s HE ERC Starting Grant No. 101040227 - WINGS. Views and opinions expressed are, however, those of the authors only and do not necessarily reflect those of the European Union or the European Research Council Executive Agency. Neither the European Union nor the granting authority can be held responsible for them. MJT acknowledges funding from the UKRI grant ST/X001075/1. ALR acknowledges funding from a Leverhulme Early Career Fellowship. We gratefully acknowledge funding from ANID - Millennium Science Initiative - AIM23-0001 and ICN12\_009 (FEB), CATA-BASAL - FB210003 (FEB), and FONDECYT Regular - 1241005 (FEB). Support for this work was also provided in part by the National Aeronautics and Space Administration through Chandra Award Numbers GO4-25074X and AR1-22010X issued by the Chandra X-ray Centre, which is operated by the Smithsonian Astrophysical Observatory for and on behalf of the National Aeronautics and Space Administration under contract NAS8-03060 and upon work supported by NASA under award 80NSSC24K0719.
\end{acknowledgements}

\bibliographystyle{aa}
\bibliography{aa55381-25}

@ARTICLE{2012ApJ...745L..34Z,
       author = {{Zubovas}, Kastytis and {King}, Andrew},
        title = "{Clearing Out a Galaxy}",
      journal = {\apjl},
         year = 2012,
        month = feb,
       volume = {745},
       number = {2},
          eid = {L34},
        pages = {L34},
          doi = {10.1088/2041-8205/745/2/L34},
archivePrefix = {arXiv},
       eprint = {1201.0866},
 primaryClass = {astro-ph.GA},
       adsurl = {https://ui.adsabs.harvard.edu/abs/2012ApJ...745L..34Z}
}

@ARTICLE{2012MNRAS.425..605F,
       author = {{Faucher-Gigu{\`e}re}, Claude-Andr{\'e} and {Quataert}, Eliot},
        title = "{The physics of galactic winds driven by active galactic nuclei}",
      journal = {\mnras},
         year = 2012,
        month = sep,
       volume = {425},
       number = {1},
        pages = {605-622},
          doi = {10.1111/j.1365-2966.2012.21512.x},
archivePrefix = {arXiv},
       eprint = {1204.2547},
 primaryClass = {astro-ph.CO},
       adsurl = {https://ui.adsabs.harvard.edu/abs/2012MNRAS.425..605F}
}

@ARTICLE{2003MNRAS.345..657K,
       author = {{King}, A.~R. and {Pounds}, K.~A.},
        title = "{Black hole winds}",
      journal = {\mnras},
         year = 2003,
        month = oct,
       volume = {345},
       number = {2},
        pages = {657-659},
          doi = {10.1046/j.1365-8711.2003.06980.x},
archivePrefix = {arXiv},
       eprint = {astro-ph/0305541},
 primaryClass = {astro-ph},
       adsurl = {https://ui.adsabs.harvard.edu/abs/2003MNRAS.345..657K}
}

@ARTICLE{2013PASJ...65...88T,
       author = {{Takeuchi}, Shun and {Ohsuga}, Ken and {Mineshige}, Shin},
        title = "{Clumpy Outflows from Supercritical Accretion Flow}",
      journal = {\pasj},
         year = 2013,
        month = aug,
       volume = {65},
          eid = {88},
        pages = {88},
          doi = {10.1093/pasj/65.4.88},
archivePrefix = {arXiv},
       eprint = {1305.1023},
 primaryClass = {astro-ph.HE},
       adsurl = {https://ui.adsabs.harvard.edu/abs/2013PASJ...65...88T}
}

@ARTICLE{2009ApJ...692..758G,
       author = {{Gibson}, Robert R. and {Jiang}, Linhua and {Brandt}, W.~N. and {Hall}, Patrick B. and {Shen}, Yue and {Wu}, Jianfeng and {Anderson}, Scott F. and {Schneider}, Donald P. and {Vanden Berk}, Daniel and {Gallagher}, S.~C. and {Fan}, Xiaohui and {York}, Donald G.},
        title = "{A Catalog of Broad Absorption Line Quasars in Sloan Digital Sky Survey Data Release 5}",
      journal = {\apj},
         year = 2009,
        month = feb,
       volume = {692},
       number = {1},
        pages = {758-777},
          doi = {10.1088/0004-637X/692/1/758},
archivePrefix = {arXiv},
       eprint = {0810.2747},
 primaryClass = {astro-ph},
       adsurl = {https://ui.adsabs.harvard.edu/abs/2009ApJ...692..758G}
}

@ARTICLE{2010A&A...521A..57T,
       author = {{Tombesi}, F. and {Cappi}, M. and {Reeves}, J.~N. and {Palumbo}, G.~G.~C. and {Yaqoob}, T. and {Braito}, V. and {Dadina}, M.},
        title = "{Evidence for ultra-fast outflows in radio-quiet AGNs. I. Detection and statistical incidence of Fe K-shell absorption lines}",
      journal = {\aap},
         year = 2010,
        month = oct,
       volume = {521},
          eid = {A57},
        pages = {A57},
          doi = {10.1051/0004-6361/200913440},
archivePrefix = {arXiv},
       eprint = {1006.2858},
 primaryClass = {astro-ph.HE},
       adsurl = {https://ui.adsabs.harvard.edu/abs/2010A&A...521A..57T}
}

@ARTICLE{2011AJ....141..167R,
       author = {{Richards}, Gordon T. and {Kruczek}, Nicholas E. and {Gallagher}, S.~C. and {Hall}, Patrick B. and {Hewett}, Paul C. and {Leighly}, Karen M. and {Deo}, Rajesh P. and {Kratzer}, Rachael M. and {Shen}, Yue},
        title = "{Unification of Luminous Type 1 Quasars through C IV Emission}",
      journal = {\aj},
         year = 2011,
        month = may,
       volume = {141},
       number = {5},
          eid = {167},
        pages = {167},
          doi = {10.1088/0004-6256/141/5/167},
archivePrefix = {arXiv},
       eprint = {1011.2282},
 primaryClass = {astro-ph.GA},
       adsurl = {https://ui.adsabs.harvard.edu/abs/2011AJ....141..167R}
}

@ARTICLE{2022ApJS..263...42W,
       author = {{Wu}, Qiaoya and {Shen}, Yue},
        title = "{A Catalog of Quasar Properties from Sloan Digital Sky Survey Data Release 16}",
      journal = {\apjs},
         year = 2022,
        month = dec,
       volume = {263},
       number = {2},
          eid = {42},
        pages = {42},
          doi = {10.3847/1538-4365/ac9ead},
archivePrefix = {arXiv},
       eprint = {2209.03987},
 primaryClass = {astro-ph.GA},
       adsurl = {https://ui.adsabs.harvard.edu/abs/2022ApJS..263...42W}
}

@ARTICLE{2020A&A...642A.150L,
       author = {{Lusso}, E. and {Risaliti}, G. and {Nardini}, E. and {Bargiacchi}, G. and {Benetti}, M. and {Bisogni}, S. and {Capozziello}, S. and {Civano}, F. and {Eggleston}, L. and {Elvis}, M. and {Fabbiano}, G. and {Gilli}, R. and {Marconi}, A. and {Paolillo}, M. and {Piedipalumbo}, E. and {Salvestrini}, F. and {Signorini}, M. and {Vignali}, C.},
        title = "{Quasars as standard candles. III. Validation of a new sample for cosmological studies}",
      journal = {\aap},
         year = 2020,
        month = oct,
       volume = {642},
          eid = {A150},
        pages = {A150},
          doi = {10.1051/0004-6361/202038899},
archivePrefix = {arXiv},
       eprint = {2008.08586},
 primaryClass = {astro-ph.GA},
       adsurl = {https://ui.adsabs.harvard.edu/abs/2020A&A...642A.150L}
}

@ARTICLE{1995ApJ...451..498M,
       author = {{Murray}, N. and {Chiang}, J. and {Grossman}, S.~A. and {Voit}, G.~M.},
        title = "{Accretion Disk Winds from Active Galactic Nuclei}",
      journal = {\apj},
         year = 1995,
        month = oct,
       volume = {451},
        pages = {498},
          doi = {10.1086/176238},
       adsurl = {https://ui.adsabs.harvard.edu/abs/1995ApJ...451..498M}
}

@ARTICLE{2019A&A...632A.109N,
       author = {{Nardini}, E. and {Lusso}, E. and {Risaliti}, G. and {Bisogni}, S. and {Civano}, F. and {Elvis}, M. and {Fabbiano}, G. and {Gilli}, R. and {Marconi}, A. and {Salvestrini}, F. and {Vignali}, C.},
        title = "{The most luminous blue quasars at 3.0 < z < 3.3. I. A tale of two X-ray populations}",
      journal = {\aap},
         year = 2019,
        month = dec,
       volume = {632},
          eid = {A109},
        pages = {A109},
          doi = {10.1051/0004-6361/201936911},
archivePrefix = {arXiv},
       eprint = {1910.04604},
 primaryClass = {astro-ph.GA},
       adsurl = {https://ui.adsabs.harvard.edu/abs/2019A&A...632A.109N}
}

@ARTICLE{2021A&A...653A.158L,
       author = {{Lusso}, E. and {Nardini}, E. and {Bisogni}, S. and {Risaliti}, G. and {Gilli}, R. and {Richards}, G.~T. and {Salvestrini}, F. and {Vignali}, C. and {Bargiacchi}, G. and {Civano}, F. and {Elvis}, M. and {Fabbiano}, G. and {Marconi}, A. and {Sacchi}, A. and {Signorini}, M.},
        title = "{The most luminous blue quasars at 3.0 < z < 3.3. II. C IV/X-ray emission and accretion disc physics}",
      journal = {\aap},
         year = 2021,
        month = sep,
       volume = {653},
          eid = {A158},
        pages = {A158},
          doi = {10.1051/0004-6361/202141356},
archivePrefix = {arXiv},
       eprint = {2107.02806},
 primaryClass = {astro-ph.GA},
       adsurl = {https://ui.adsabs.harvard.edu/abs/2021A&A...653A.158L}
}

@ARTICLE{2000ApJ...543..686P,
       author = {{Proga}, Daniel and {Stone}, James M. and {Kallman}, Timothy R.},
        title = "{Dynamics of Line-driven Disk Winds in Active Galactic Nuclei}",
      journal = {\apj},
         year = 2000,
        month = nov,
       volume = {543},
       number = {2},
        pages = {686-696},
          doi = {10.1086/317154},
archivePrefix = {arXiv},
       eprint = {astro-ph/0005315},
 primaryClass = {astro-ph},
       adsurl = {https://ui.adsabs.harvard.edu/abs/2000ApJ...543..686P}
}

@ARTICLE{2018MNRAS.480.5184N,
       author = {{Ni}, Q. and {Brandt}, W.~N. and {Luo}, B. and {Hall}, P.~B. and {Shen}, Yue and {Anderson}, S.~F. and {Plotkin}, R.~M. and {Richards}, Gordon T. and {Schneider}, D.~P. and {Shemmer}, O. and {Wu}, Jianfeng},
        title = "{Connecting the X-ray properties of weak-line and typical quasars: testing for a geometrically thick accretion disk}",
      journal = {\mnras},
         year = 2018,
        month = nov,
       volume = {480},
       number = {4},
        pages = {5184-5202},
          doi = {10.1093/mnras/sty1989},
archivePrefix = {arXiv},
       eprint = {1807.08757},
 primaryClass = {astro-ph.GA},
       adsurl = {https://ui.adsabs.harvard.edu/abs/2018MNRAS.480.5184N}
}

@ARTICLE{2004ApJ...616..688P,
       author = {{Proga}, Daniel and {Kallman}, Timothy R.},
        title = "{Dynamics of Line-driven Disk Winds in Active Galactic Nuclei. II. Effects of Disk Radiation}",
      journal = {\apj},
         year = 2004,
        month = dec,
       volume = {616},
       number = {2},
        pages = {688-695},
          doi = {10.1086/425117},
archivePrefix = {arXiv},
       eprint = {astro-ph/0408293},
 primaryClass = {astro-ph},
       adsurl = {https://ui.adsabs.harvard.edu/abs/2004ApJ...616..688P}
}

@ARTICLE{2020A&A...635L...5Z,
       author = {{Zappacosta}, L. and {Piconcelli}, E. and {Giustini}, M. and {Vietri}, G. and {Duras}, F. and {Miniutti}, G. and {Bischetti}, M. and {Bongiorno}, A. and {Brusa}, M. and {Chiaberge}, M. and {Comastri}, A. and {Feruglio}, C. and {Luminari}, A. and {Marconi}, A. and {Ricci}, C. and {Vignali}, C. and {Fiore}, F.},
        title = "{The WISSH quasars project. VII. The impact of extreme radiative field in the accretion disc and X-ray corona interplay}",
      journal = {\aap},
         year = 2020,
        month = mar,
       volume = {635},
          eid = {L5},
        pages = {L5},
          doi = {10.1051/0004-6361/201937292},
archivePrefix = {arXiv},
       eprint = {2002.00957},
 primaryClass = {astro-ph.GA},
       adsurl = {https://ui.adsabs.harvard.edu/abs/2020A&A...635L...5Z}
}

@ARTICLE{2023A&A...677A.111T,
       author = {{Trefoloni}, Bartolomeo and {Lusso}, Elisabeta and {Nardini}, Emanuele and {Risaliti}, Guido and {Bargiacchi}, Giada and {Bisogni}, Susanna and {Civano}, Francesca M. and {Elvis}, Martin and {Fabbiano}, Giuseppina and {Gilli}, Roberto and {Marconi}, Alessandro and {Richards}, Gordon T. and {Sacchi}, Andrea and {Salvestrini}, Francesco and {Signorini}, Matilde and {Vignali}, Cristian},
        title = "{The most luminous blue quasars at 3.0 < z < 3.3. III. LBT spectra and accretion parameters}",
      journal = {\aap},
         year = 2023,
        month = sep,
       volume = {677},
          eid = {A111},
        pages = {A111},
          doi = {10.1051/0004-6361/202346024},
archivePrefix = {arXiv},
       eprint = {2305.07699},
 primaryClass = {astro-ph.GA},
       adsurl = {https://ui.adsabs.harvard.edu/abs/2023A&A...677A.111T}
}

@ARTICLE{2008ApJ...685..773G,
       author = {{Gibson}, Robert R. and {Brandt}, W.~N. and {Schneider}, Donald P.},
        title = "{Are Optically Selected Quasars Universally X-Ray Luminous? X-Ray-UV Relations in Sloan Digital Sky Survey Quasars}",
      journal = {\apj},
         year = 2008,
        month = oct,
       volume = {685},
       number = {2},
        pages = {773-786},
          doi = {10.1086/590403},
archivePrefix = {arXiv},
       eprint = {0808.2603},
 primaryClass = {astro-ph},
       adsurl = {https://ui.adsabs.harvard.edu/abs/2008ApJ...685..773G}
}

@ARTICLE{2006ApJ...644..709G,
       author = {{Gallagher}, S.~C. and {Brandt}, W.~N. and {Chartas}, G. and {Priddey}, R. and {Garmire}, G.~P. and {Sambruna}, R.~M.},
        title = "{An Exploratory Chandra Survey of a Well-defined Sample of 35 Large Bright Quasar Survey Broad Absorption Line Quasars}",
      journal = {\apj},
         year = 2006,
        month = jun,
       volume = {644},
       number = {2},
        pages = {709-724},
          doi = {10.1086/503762},
archivePrefix = {arXiv},
       eprint = {astro-ph/0602550},
 primaryClass = {astro-ph},
       adsurl = {https://ui.adsabs.harvard.edu/abs/2006ApJ...644..709G}
}

@ARTICLE{2002ApJ...567...37G,
       author = {{Gallagher}, S.~C. and {Brandt}, W.~N. and {Chartas}, G. and {Garmire}, G.~P.},
        title = "{X-Ray Spectroscopy of Quasi-Stellar Objects with Broad Ultraviolet Absorption Lines}",
      journal = {\apj},
         year = 2002,
        month = mar,
       volume = {567},
       number = {1},
        pages = {37-41},
          doi = {10.1086/338485},
archivePrefix = {arXiv},
       eprint = {astro-ph/0110579},
 primaryClass = {astro-ph},
       adsurl = {https://ui.adsabs.harvard.edu/abs/2002ApJ...567...37G}
}

@ARTICLE{2011MNRAS.413.1013S,
       author = {{Stalin}, C.~S. and {Srianand}, R. and {Petitjean}, P.},
        title = "{X-ray and optical properties of broad absorption line quasars in the Canada-France-Hawaii Telescope Legacy Survey}",
      journal = {\mnras},
         year = 2011,
        month = may,
       volume = {413},
       number = {2},
        pages = {1013-1023},
          doi = {10.1111/j.1365-2966.2010.18190.x},
archivePrefix = {arXiv},
       eprint = {1012.2425},
 primaryClass = {astro-ph.CO},
       adsurl = {https://ui.adsabs.harvard.edu/abs/2011MNRAS.413.1013S}
}

@ARTICLE{2015ApJ...805..122L,
       author = {{Luo}, B. and {Brandt}, W.~N. and {Hall}, P.~B. and {Wu}, Jianfeng and {Anderson}, S.~F. and {Garmire}, G.~P. and {Gibson}, R.~R. and {Plotkin}, R.~M. and {Richards}, G.~T. and {Schneider}, D.~P. and {Shemmer}, O. and {Shen}, Yue},
        title = "{X-ray Insights into the Nature of PHL 1811 Analogs and Weak Emission-line Quasars: Unification with a Geometrically Thick Accretion Disk?}",
      journal = {\apj},
         year = 2015,
        month = jun,
       volume = {805},
       number = {2},
          eid = {122},
        pages = {122},
          doi = {10.1088/0004-637X/805/2/122},
archivePrefix = {arXiv},
       eprint = {1503.02085},
 primaryClass = {astro-ph.GA},
       adsurl = {https://ui.adsabs.harvard.edu/abs/2015ApJ...805..122L}
}

@ARTICLE{2020ApJS..250....8L,
       author = {{Lyke}, Brad W. and {Higley}, Alexandra N. and {McLane}, J.~N. and {Schurhammer}, Danielle P. and {Myers}, Adam D. and {Ross}, Ashley J. and {Dawson}, Kyle and {Chabanier}, Sol{\`e}ne and {Martini}, Paul and {Busca}, Nicol{\'a}s G. and {Mas des Bourboux}, H{\'e}lion du and {Salvato}, Mara and {Streblyanska}, Alina and {Zarrouk}, Pauline and {Burtin}, Etienne and {Anderson}, Scott F. and {Bautista}, Julian and {Bizyaev}, Dmitry and {Brandt}, W.~N. and {Brinkmann}, Jonathan and {Brownstein}, Joel R. and {Comparat}, Johan and {Green}, Paul and {de la Macorra}, Axel and {Mu{\~n}oz Guti{\'e}rrez}, Andrea and {Hou}, Jiamin and {Newman}, Jeffrey A. and {Palanque-Delabrouille}, Nathalie and {P{\^a}ris}, Isabelle and {Percival}, Will J. and {Petitjean}, Patrick and {Rich}, James and {Rossi}, Graziano and {Schneider}, Donald P. and {Smith}, Alexander and {Vivek}, M. and {Weaver}, Benjamin Alan},
        title = "{The Sloan Digital Sky Survey Quasar Catalog: Sixteenth Data Release}",
      journal = {\apjs},
         year = 2020,
        month = sep,
       volume = {250},
       number = {1},
          eid = {8},
        pages = {8},
          doi = {10.3847/1538-4365/aba623},
archivePrefix = {arXiv},
       eprint = {2007.09001},
 primaryClass = {astro-ph.GA},
       adsurl = {https://ui.adsabs.harvard.edu/abs/2020ApJS..250....8L}
}

@ARTICLE{2013ascl.soft11006C,
       author = {{CIAO Development Team}},
        title = "{CIAO: Chandra Interactive Analysis of Observations}",
      journal = {ASCL},
         year = 2013,
        month = nov,
          eid = {ascl:1311.006},
       adsurl = {https://ui.adsabs.harvard.edu/abs/2013ascl.soft11006C}
}

@ARTICLE{1999ascl.soft10005A,
       author = {{Arnaud}, Keith and {Dorman}, Ben and {Gordon}, Craig},
        title = "{XSPEC: An X-ray spectral fitting package}",
      journal = {ASCL},
         year = 1999,
        month = oct,
          eid = {ascl:9910.005},
       adsurl = {https://ui.adsabs.harvard.edu/abs/1999ascl.soft10005A}
}

@ARTICLE{1979ApJ...228..939C,
       author = {{Cash}, W.},
        title = "{Parameter estimation in astronomy through application of the likelihood ratio.}",
      journal = {\apj},
         year = 1979,
        month = mar,
       volume = {228},
        pages = {939-947},
          doi = {10.1086/156922},
       adsurl = {https://ui.adsabs.harvard.edu/abs/1979ApJ...228..939C}
}

@ARTICLE{2017A&A...605A..51K,
       author = {{Kaastra}, J.~S.},
        title = "{On the use of C-stat in testing models for X-ray spectra}",
      journal = {\aap},
         year = 2017,
        month = sep,
       volume = {605},
          eid = {A51},
        pages = {A51},
          doi = {10.1051/0004-6361/201629319},
archivePrefix = {arXiv},
       eprint = {1707.09202},
 primaryClass = {astro-ph.HE},
       adsurl = {https://ui.adsabs.harvard.edu/abs/2017A&A...605A..51K}
}

@ARTICLE{2016A&A...594A.116H,
       author = {{HI4PI Collaboration}},
        title = "{HI4PI: A full-sky H I survey based on EBHIS and GASS}",
      journal = {\aap},
         year = 2016,
        month = oct,
       volume = {594},
          eid = {A116},
        pages = {A116},
          doi = {10.1051/0004-6361/201629178},
archivePrefix = {arXiv},
       eprint = {1610.06175},
 primaryClass = {astro-ph.GA},
       adsurl = {https://ui.adsabs.harvard.edu/abs/2016A&A...594A.116H}
}

@ARTICLE{2015ApJ...815...33R,
       author = {{Risaliti}, G. and {Lusso}, E.},
        title = "{A Hubble Diagram for Quasars}",
      journal = {\apj},
         year = 2015,
        month = dec,
       volume = {815},
       number = {1},
          eid = {33},
        pages = {33},
          doi = {10.1088/0004-637X/815/1/33},
archivePrefix = {arXiv},
       eprint = {1505.07118},
 primaryClass = {astro-ph.CO},
       adsurl = {https://ui.adsabs.harvard.edu/abs/2015ApJ...815...33R}
}

@ARTICLE{2020A&A...641A.137T,
       author = {{Traulsen}, I. and {Schwope}, A.~D. and {Lamer}, G. and {Ballet}, J. and {Carrera}, F.~J. and {Ceballos}, M.~T. and {Coriat}, M. and {Freyberg}, M.~J. and {Koliopanos}, F. and {Kurpas}, J. and {Michel}, L. and {Motch}, C. and {Page}, M.~J. and {Watson}, M.~G. and {Webb}, N.~A.},
        title = "{The XMM-Newton serendipitous survey. X. The second source catalogue from overlapping XMM-Newton observations and its long-term variable content}",
      journal = {\aap},
         year = 2020,
        month = sep,
       volume = {641},
          eid = {A137},
        pages = {A137},
          doi = {10.1051/0004-6361/202037706},
archivePrefix = {arXiv},
       eprint = {2007.02932},
 primaryClass = {astro-ph.HE},
       adsurl = {https://ui.adsabs.harvard.edu/abs/2020A&A...641A.137T}
}

@ARTICLE{2013ascl.soft03002F,
       author = {{Foreman-Mackey}, Daniel and {Conley}, Alex and {Meierjurgen Farr}, Will and {Hogg}, David W. and {Lang}, Dustin and {Marshall}, Phil and {Price-Whelan}, Adrian and {Sanders}, Jeremy and {Zuntz}, Joe},
        title = "{emcee: The MCMC Hammer}",
      journal = {ASCL},
         year = 2013,
        month = mar,
          eid = {ascl:1303.002},
       adsurl = {https://ui.adsabs.harvard.edu/abs/2013ascl.soft03002F}
}

@ARTICLE{1998AJ....115.2285M,
       author = {{Magorrian}, John and {Tremaine}, Scott and {Richstone}, Douglas and {Bender}, Ralf and {Bower}, Gary and {Dressler}, Alan and {Faber}, S.~M. and {Gebhardt}, Karl and {Green}, Richard and {Grillmair}, Carl and {Kormendy}, John and {Lauer}, Tod},
        title = "{The Demography of Massive Dark Objects in Galaxy Centers}",
      journal = {\aj},
         year = 1998,
        month = jun,
       volume = {115},
       number = {6},
        pages = {2285-2305},
          doi = {10.1086/300353},
archivePrefix = {arXiv},
       eprint = {astro-ph/9708072},
 primaryClass = {astro-ph},
       adsurl = {https://ui.adsabs.harvard.edu/abs/1998AJ....115.2285M}
}

@ARTICLE{2000ApJ...539L...9F,
       author = {{Ferrarese}, Laura and {Merritt}, David},
        title = "{A Fundamental Relation between Supermassive Black Holes and Their Host Galaxies}",
      journal = {\apjl},
         year = 2000,
        month = aug,
       volume = {539},
       number = {1},
        pages = {L9-L12},
          doi = {10.1086/312838},
archivePrefix = {arXiv},
       eprint = {astro-ph/0006053},
 primaryClass = {astro-ph},
       adsurl = {https://ui.adsabs.harvard.edu/abs/2000ApJ...539L...9F}
}

@ARTICLE{2013ARA&A..51..511K,
       author = {{Kormendy}, John and {Ho}, Luis C.},
        title = "{Coevolution (Or Not) of Supermassive Black Holes and Host Galaxies}",
      journal = {\araa},
         year = 2013,
        month = aug,
       volume = {51},
       number = {1},
        pages = {511-653},
          doi = {10.1146/annurev-astro-082708-101811},
archivePrefix = {arXiv},
       eprint = {1304.7762},
 primaryClass = {astro-ph.CO},
       adsurl = {https://ui.adsabs.harvard.edu/abs/2013ARA&A..51..511K}
}

@ARTICLE{2000ApJ...539L..13G,
       author = {{Gebhardt}, Karl and {Bender}, Ralf and {Bower}, Gary and {Dressler}, Alan and {Faber}, S.~M. and {Filippenko}, Alexei V. and {Green}, Richard and {Grillmair}, Carl and {Ho}, Luis C. and {Kormendy}, John and {Lauer}, Tod R. and {Magorrian}, John and {Pinkney}, Jason and {Richstone}, Douglas and {Tremaine}, Scott},
        title = "{A Relationship between Nuclear Black Hole Mass and Galaxy Velocity Dispersion}",
      journal = {\apjl},
         year = 2000,
        month = aug,
       volume = {539},
       number = {1},
        pages = {L13-L16},
          doi = {10.1086/312840},
archivePrefix = {arXiv},
       eprint = {astro-ph/0006289},
 primaryClass = {astro-ph},
       adsurl = {https://ui.adsabs.harvard.edu/abs/2000ApJ...539L..13G}
}

@ARTICLE{2017NatAs...1E.165H,
       author = {{Harrison}, C.~M.},
        title = "{Impact of supermassive black hole growth on star formation}",
      journal = {Nat. Astron.},
         year = 2017,
        month = jul,
       volume = {1},
          eid = {0165},
        pages = {0165},
          doi = {10.1038/s41550-017-0165},
archivePrefix = {arXiv},
       eprint = {1703.06889},
 primaryClass = {astro-ph.GA},
       adsurl = {https://ui.adsabs.harvard.edu/abs/2017NatAs...1E.165H}
}

@ARTICLE{2020ApJ...894...28D,
       author = {{Davies}, Rebecca L. and {F{\"o}rster Schreiber}, N.~M. and {Lutz}, D. and {Genzel}, R. and {Belli}, S. and {Shimizu}, T.~T. and {Contursi}, A. and {Davies}, R.~I. and {Herrera-Camus}, R. and {Lee}, M.~M. and {Naab}, T. and {Price}, S.~H. and {Renzini}, A. and {Schruba}, A. and {Sternberg}, A. and {Tacconi}, L.~J. and {{\"U}bler}, H. and {Wisnioski}, E. and {Wuyts}, S.},
        title = "{From Nuclear to Circumgalactic: Zooming in on AGN-driven Outflows at z {\ensuremath{\sim}} 2.2 with SINFONI}",
      journal = {\apj},
         year = 2020,
        month = may,
       volume = {894},
       number = {1},
          eid = {28},
        pages = {28},
          doi = {10.3847/1538-4357/ab86ad},
archivePrefix = {arXiv},
       eprint = {2004.02891},
 primaryClass = {astro-ph.GA},
       adsurl = {https://ui.adsabs.harvard.edu/abs/2020ApJ...894...28D}
}

@ARTICLE{2024MNRAS.528.4976D,
       author = {{Davies}, Rebecca L. and {Belli}, Sirio and {Park}, Minjung and {Mendel}, J. Trevor and {Johnson}, Benjamin D. and {Conroy}, Charlie and {Benton}, Chlo{\"e} and {Bugiani}, Letizia and {Emami}, Razieh and {Leja}, Joel and {Li}, Yijia and {Maheson}, Gabriel and {Mathews}, Elijah P. and {Naidu}, Rohan P. and {Nelson}, Erica J. and {Tacchella}, Sandro and {Terrazas}, Bryan A. and {Weinberger}, Rainer},
        title = "{JWST reveals widespread AGN-driven neutral gas outflows in massive z   2 galaxies}",
      journal = {\mnras},
         year = 2024,
        month = mar,
       volume = {528},
       number = {3},
        pages = {4976-4992},
          doi = {10.1093/mnras/stae327},
archivePrefix = {arXiv},
       eprint = {2310.17939},
 primaryClass = {astro-ph.GA},
       adsurl = {https://ui.adsabs.harvard.edu/abs/2024MNRAS.528.4976D}
}

@ARTICLE{2012ARA&A..50..455F,
       author = {{Fabian}, A.~C.},
        title = "{Observational Evidence of Active Galactic Nuclei Feedback}",
      journal = {\araa},
         year = 2012,
        month = sep,
       volume = {50},
        pages = {455-489},
          doi = {10.1146/annurev-astro-081811-125521},
archivePrefix = {arXiv},
       eprint = {1204.4114},
 primaryClass = {astro-ph.CO},
       adsurl = {https://ui.adsabs.harvard.edu/abs/2012ARA&A..50..455F}
}

@ARTICLE{2003AJ....125.1784H,
       author = {{Hewett}, Paul C. and {Foltz}, Craig B.},
        title = "{The Frequency and Radio Properties of Broad Absorption Line Quasars}",
      journal = {\aj},
         year = 2003,
        month = apr,
       volume = {125},
       number = {4},
        pages = {1784-1794},
          doi = {10.1086/368392},
archivePrefix = {arXiv},
       eprint = {astro-ph/0301191},
 primaryClass = {astro-ph},
       adsurl = {https://ui.adsabs.harvard.edu/abs/2003AJ....125.1784H}
}

@ARTICLE{2020MNRAS.492..719T,
       author = {{Timlin}, John D. and {Brandt}, W.~N. and {Ni}, Q. and {Luo}, B. and {Pu}, Xingting and {Schneider}, D.~P. and {Vivek}, M. and {Yi}, W.},
        title = "{The correlations between optical/UV broad lines and X-ray emission for a large sample of quasars}",
      journal = {\mnras},
         year = 2020,
        month = feb,
       volume = {492},
       number = {1},
        pages = {719-741},
          doi = {10.1093/mnras/stz3433},
archivePrefix = {arXiv},
       eprint = {1912.02189},
 primaryClass = {astro-ph.GA},
       adsurl = {https://ui.adsabs.harvard.edu/abs/2020MNRAS.492..719T}
}

@ARTICLE{2013MNRAS.430...60G,
       author = {{Gofford}, Jason and {Reeves}, James N. and {Tombesi}, Francesco and {Braito}, Valentina and {Turner}, T. Jane and {Miller}, Lance and {Cappi}, Massimo},
        title = "{The Suzaku view of highly ionized outflows in AGN - I. Statistical detection and global absorber properties}",
      journal = {\mnras},
         year = 2013,
        month = mar,
       volume = {430},
       number = {1},
        pages = {60-80},
          doi = {10.1093/mnras/sts481},
archivePrefix = {arXiv},
       eprint = {1211.5810},
 primaryClass = {astro-ph.HE},
       adsurl = {https://ui.adsabs.harvard.edu/abs/2013MNRAS.430...60G}
}

@ARTICLE{2012A&A...537L...8C,
       author = {{Cano-D{\'\i}az}, M. and {Maiolino}, R. and {Marconi}, A. and {Netzer}, H. and {Shemmer}, O. and {Cresci}, G.},
        title = "{Observational evidence of quasar feedback quenching star formation at high redshift}",
      journal = {\aap},
         year = 2012,
        month = jan,
       volume = {537},
          eid = {L8},
        pages = {L8},
          doi = {10.1051/0004-6361/201118358},
archivePrefix = {arXiv},
       eprint = {1112.3071},
 primaryClass = {astro-ph.CO},
       adsurl = {https://ui.adsabs.harvard.edu/abs/2012A&A...537L...8C}
}

@ARTICLE{2015ARA&A..53..115K,
       author = {{King}, Andrew and {Pounds}, Ken},
        title = "{Powerful Outflows and Feedback from Active Galactic Nuclei}",
      journal = {\araa},
         year = 2015,
        month = aug,
       volume = {53},
        pages = {115-154},
          doi = {10.1146/annurev-astro-082214-122316},
archivePrefix = {arXiv},
       eprint = {1503.05206},
 primaryClass = {astro-ph.GA},
       adsurl = {https://ui.adsabs.harvard.edu/abs/2015ARA&A..53..115K}
}

@ARTICLE{2002ApJ...574..740T,
       author = {{Tremaine}, Scott and {Gebhardt}, Karl and {Bender}, Ralf and {Bower}, Gary and {Dressler}, Alan and {Faber}, S.~M. and {Filippenko}, Alexei V. and {Green}, Richard and {Grillmair}, Carl and {Ho}, Luis C. and {Kormendy}, John and {Lauer}, Tod R. and {Magorrian}, John and {Pinkney}, Jason and {Richstone}, Douglas},
        title = "{The Slope of the Black Hole Mass versus Velocity Dispersion Correlation}",
      journal = {\apj},
         year = 2002,
        month = aug,
       volume = {574},
       number = {2},
        pages = {740-753},
          doi = {10.1086/341002},
archivePrefix = {arXiv},
       eprint = {astro-ph/0203468},
 primaryClass = {astro-ph},
       adsurl = {https://ui.adsabs.harvard.edu/abs/2002ApJ...574..740T}
}

@ARTICLE{2004ApJ...604L..89H,
       author = {{H{\"a}ring}, Nadine and {Rix}, Hans-Walter},
        title = "{On the Black Hole Mass-Bulge Mass Relation}",
      journal = {\apjl},
         year = 2004,
        month = apr,
       volume = {604},
       number = {2},
        pages = {L89-L92},
          doi = {10.1086/383567},
archivePrefix = {arXiv},
       eprint = {astro-ph/0402376},
 primaryClass = {astro-ph},
       adsurl = {https://ui.adsabs.harvard.edu/abs/2004ApJ...604L..89H}
}

@ARTICLE{2006ApJS..165....1T,
       author = {{Trump}, Jonathan R. and {Hall}, Patrick B. and {Reichard}, Timothy A. and {Richards}, Gordon T. and {Schneider}, Donald P. and {Vanden Berk}, Daniel E. and {Knapp}, Gillian R. and {Anderson}, Scott F. and {Fan}, Xiaohui and {Brinkman}, J. and {Kleinman}, S.~J. and {Nitta}, Atsuko},
        title = "{A Catalog of Broad Absorption Line Quasars from the Sloan Digital Sky Survey Third Data Release}",
      journal = {\apjs},
         year = 2006,
        month = jul,
       volume = {165},
       number = {1},
        pages = {1-18},
          doi = {10.1086/503834},
archivePrefix = {arXiv},
       eprint = {astro-ph/0603070},
 primaryClass = {astro-ph},
       adsurl = {https://ui.adsabs.harvard.edu/abs/2006ApJS..165....1T}
}

@ARTICLE{2009ASPC..411..251M,
       author = {{Markwardt}, C.~B.},
        title = "{Non-linear Least-squares Fitting in IDL with MPFIT}",
         year = 2009,
       journal = {ASP Conf. Ser.},
       volume = {411},
        month = sep,
        pages = {251},
          doi = {10.48550/arXiv.0902.2850},
archivePrefix = {arXiv},
       eprint = {0902.2850},
 primaryClass = {astro-ph.IM},
       adsurl = {https://ui.adsabs.harvard.edu/abs/2009ASPC..411..251M}
}

@ARTICLE{1978LNM...630..105M,
       author = {{Mor{\'e}}, Jorge J.},
        title = "{The Levenberg-Marquardt algorithm: Implementation and theory}",
    journal = {Numerical Analysis (Springer)},
         year = 1978,
       volume = {630},
        pages = {105-116},
          doi = {10.1007/BFb0067700},
       adsurl = {https://ui.adsabs.harvard.edu/abs/1978LNM...630..105M}
}

@ARTICLE{2023MNRAS.523..646T,
       author = {{Temple}, Matthew J. and {Matthews}, James H. and {Hewett}, Paul C. and {Rankine}, Amy L. and {Richards}, Gordon T. and {Banerji}, Manda and {Ferland}, Gary J. and {Knigge}, Christian and {Stepney}, Matthew},
        title = "{Testing AGN outflow and accretion models with C IV and He II emission line demographics in z {\ensuremath{\approx}} 2 quasars}",
      journal = {\mnras},
         year = 2023,
        month = jul,
       volume = {523},
       number = {1},
        pages = {646-666},
          doi = {10.1093/mnras/stad1448},
archivePrefix = {arXiv},
       eprint = {2301.02675},
 primaryClass = {astro-ph.GA},
       adsurl = {https://ui.adsabs.harvard.edu/abs/2023MNRAS.523..646T}
}

@ARTICLE{2020MNRAS.492.4553R,
       author = {{Rankine}, Amy L. and {Hewett}, Paul C. and {Banerji}, Manda and {Richards}, Gordon T.},
        title = "{BAL and non-BAL quasars: continuum, emission, and absorption properties establish a common parent sample}",
      journal = {\mnras},
         year = 2020,
        month = mar,
       volume = {492},
       number = {3},
        pages = {4553-4575},
          doi = {10.1093/mnras/staa130},
archivePrefix = {arXiv},
       eprint = {1912.08700},
 primaryClass = {astro-ph.GA},
       adsurl = {https://ui.adsabs.harvard.edu/abs/2020MNRAS.492.4553R}
}

@ARTICLE{2023ApJ...952...44B,
       author = {{Bischetti}, Manuela and {Fiore}, Fabrizio and {Feruglio}, Chiara and {D'Odorico}, Valentina and {Arav}, Nahum and {Costa}, Tiago and {Zubovas}, Kastytis and {Becker}, George and {Bosman}, Sarah E.~I. and {Cupani}, Guido and {Davies}, Rebecca and {Eilers}, Anna-Christina and {Farina}, Emanuele Paolo and {Ferrara}, Andrea and {Gaspari}, Massimo and {Mazzucchelli}, Chiara and {Onoue}, Masafusa and {Piconcelli}, Enrico and {Zanchettin}, Maria Vittoria and {Zhu}, Yongda},
        title = "{The Fraction and Kinematics of Broad Absorption Line Quasars across Cosmic Time}",
      journal = {\apj},
         year = 2023,
        month = jul,
       volume = {952},
       number = {1},
          eid = {44},
        pages = {44},
          doi = {10.3847/1538-4357/accea4},
archivePrefix = {arXiv},
       eprint = {2301.09731},
 primaryClass = {astro-ph.GA},
       adsurl = {https://ui.adsabs.harvard.edu/abs/2023ApJ...952...44B}
}

@ARTICLE{2016MNRAS.461..647C,
       author = {{Coatman}, Liam and {Hewett}, Paul C. and {Banerji}, Manda and {Richards}, Gordon T.},
        title = "{C IV emission-line properties and systematic trends in quasar black hole mass estimates}",
      journal = {\mnras},
         year = 2016,
        month = sep,
       volume = {461},
       number = {1},
        pages = {647-665},
          doi = {10.1093/mnras/stw1360},
archivePrefix = {arXiv},
       eprint = {1606.02726},
 primaryClass = {astro-ph.GA},
       adsurl = {https://ui.adsabs.harvard.edu/abs/2016MNRAS.461..647C}
}

@ARTICLE{2019A&A...628L...4L,
       author = {{Lusso}, E. and {Piedipalumbo}, E. and {Risaliti}, G. and {Paolillo}, M. and {Bisogni}, S. and {Nardini}, E. and {Amati}, L.},
        title = "{Tension with the flat {\ensuremath{\Lambda}}CDM model from a high-redshift Hubble diagram of supernovae, quasars, and gamma-ray bursts}",
      journal = {\aap},
         year = 2019,
        month = aug,
       volume = {628},
          eid = {L4},
        pages = {L4},
          doi = {10.1051/0004-6361/201936223},
archivePrefix = {arXiv},
       eprint = {1907.07692},
 primaryClass = {astro-ph.CO},
       adsurl = {https://ui.adsabs.harvard.edu/abs/2019A&A...628L...4L}
}

@ARTICLE{2019NatAs...3..272R,
       author = {{Risaliti}, G. and {Lusso}, E.},
        title = "{Cosmological Constraints from the Hubble Diagram of Quasars at High Redshifts}",
      journal = {Nat. Astron.},
         year = 2019,
        month = jan,
       volume = {3},
        pages = {272-277},
          doi = {10.1038/s41550-018-0657-z},
archivePrefix = {arXiv},
       eprint = {1811.02590},
 primaryClass = {astro-ph.CO},
       adsurl = {https://ui.adsabs.harvard.edu/abs/2019NatAs...3..272R}
}

@ARTICLE{2013RMxAA..49..137F,
       author = {{Ferland}, G.~J. and {Porter}, R.~L. and {van Hoof}, P.~A.~M. and {Williams}, R.~J.~R. and {Abel}, N.~P. and {Lykins}, M.~L. and {Shaw}, G. and {Henney}, W.~J. and {Stancil}, P.~C.},
        title = "{The 2013 Release of Cloudy}",
      journal = {Rev. Mex. Astron. Astrofis.},
         year = 2013,
        month = apr,
       volume = {49},
        pages = {137-163},
          doi = {10.48550/arXiv.1302.4485},
archivePrefix = {arXiv},
       eprint = {1302.4485},
 primaryClass = {astro-ph.GA},
       adsurl = {https://ui.adsabs.harvard.edu/abs/2013RMxAA..49..137F}
}

@ARTICLE{1990ARA&A..28..215D,
       author = {{Dickey}, John M. and {Lockman}, Felix J.},
        title = "{H I in the galaxy.}",
      journal = {\araa},
         year = 1990,
        month = jan,
       volume = {28},
        pages = {215-261},
          doi = {10.1146/annurev.aa.28.090190.001243},
       adsurl = {https://ui.adsabs.harvard.edu/abs/1990ARA&A..28..215D}
}

@ARTICLE{2005A&A...440..775K,
       author = {{Kalberla}, P.~M.~W. and {Burton}, W.~B. and {Hartmann}, Dap and {Arnal}, E.~M. and {Bajaja}, E. and {Morras}, R. and {P{\"o}ppel}, W.~G.~L.},
        title = "{The Leiden/Argentine/Bonn (LAB) Survey of Galactic HI. Final data release of the combined LDS and IAR surveys with improved stray-radiation corrections}",
      journal = {\aap},
         year = 2005,
        month = sep,
       volume = {440},
       number = {2},
        pages = {775-782},
          doi = {10.1051/0004-6361:20041864},
archivePrefix = {arXiv},
       eprint = {astro-ph/0504140},
 primaryClass = {astro-ph},
       adsurl = {https://ui.adsabs.harvard.edu/abs/2005A&A...440..775K}
}

@ARTICLE{2023MNRAS.526.3967M,
       author = {{Matthews}, James H. and {Strong-Wright}, Jago and {Knigge}, Christian and {Hewett}, Paul and {Temple}, Matthew J. and {Long}, Knox S. and {Rankine}, Amy L. and {Stepney}, Matthew and {Banerji}, Manda and {Richards}, Gordon T.},
        title = "{A disc wind model for blueshifts in quasar broad emission lines}",
      journal = {\mnras},
         year = 2023,
        month = dec,
       volume = {526},
       number = {3},
        pages = {3967-3986},
          doi = {10.1093/mnras/stad2895},
archivePrefix = {arXiv},
       eprint = {2309.14434},
 primaryClass = {astro-ph.GA},
       adsurl = {https://ui.adsabs.harvard.edu/abs/2023MNRAS.526.3967M}
}

@ARTICLE{2011MNRAS.410..860A,
       author = {{Allen}, James T. and {Hewett}, Paul C. and {Maddox}, Natasha and {Richards}, Gordon T. and {Belokurov}, Vasily},
        title = "{A strong redshift dependence of the broad absorption line quasar fraction}",
      journal = {\mnras},
         year = 2011,
        month = jan,
       volume = {410},
       number = {2},
        pages = {860-884},
          doi = {10.1111/j.1365-2966.2010.17489.x},
archivePrefix = {arXiv},
       eprint = {1007.3991},
 primaryClass = {astro-ph.CO},
       adsurl = {https://ui.adsabs.harvard.edu/abs/2011MNRAS.410..860A}
}

@ARTICLE{2014ApJ...789...19H,
       author = {{Higginbottom}, Nick and {Proga}, Daniel and {Knigge}, Christian and {Long}, Knox S. and {Matthews}, James H. and {Sim}, Stuart A.},
        title = "{Line-driven Disk Winds in Active Galactic Nuclei: The Critical Importance of Ionization and Radiative Transfer}",
      journal = {\apj},
         year = 2014,
        month = jul,
       volume = {789},
       number = {1},
          eid = {19},
        pages = {19},
          doi = {10.1088/0004-637X/789/1/19},
archivePrefix = {arXiv},
       eprint = {1402.1849},
 primaryClass = {astro-ph.GA},
       adsurl = {https://ui.adsabs.harvard.edu/abs/2014ApJ...789...19H}
}

@ARTICLE{2024MNRAS.527.9236H,
       author = {{Higginbottom}, Nick and {Scepi}, Nicolas and {Knigge}, Christian and {Long}, Knox S. and {Matthews}, James H. and {Sim}, Stuart A.},
        title = "{State-of-the-art simulations of line-driven accretion disc winds: realistic radiation hydrodynamics leads to weaker outflows}",
      journal = {\mnras},
         year = 2024,
        month = jan,
       volume = {527},
       number = {3},
        pages = {9236-9249},
          doi = {10.1093/mnras/stad3830},
archivePrefix = {arXiv},
       eprint = {2312.06042},
 primaryClass = {astro-ph.HE},
       adsurl = {https://ui.adsabs.harvard.edu/abs/2024MNRAS.527.9236H}
}

@ARTICLE{2024MNRAS.530.5143D,
       author = {{Dyda}, Sergei and {Davis}, Shane W. and {Proga}, Daniel},
        title = "{Time-dependent AGN disc winds - I. X-ray irradiation}",
      journal = {\mnras},
         year = 2024,
        month = jun,
       volume = {530},
       number = {4},
        pages = {5143-5154},
          doi = {10.1093/mnras/stae1159},
archivePrefix = {arXiv},
       eprint = {2310.18557},
 primaryClass = {astro-ph.HE},
       adsurl = {https://ui.adsabs.harvard.edu/abs/2024MNRAS.530.5143D}
}

@ARTICLE{2017A&A...608A..51M,
       author = {{Martocchia}, S. and {Piconcelli}, E. and {Zappacosta}, L. and {Duras}, F. and {Vietri}, G. and {Vignali}, C. and {Bianchi}, S. and {Bischetti}, M. and {Bongiorno}, A. and {Brusa}, M. and {Lanzuisi}, G. and {Marconi}, A. and {Mathur}, S. and {Miniutti}, G. and {Nicastro}, F. and {Bruni}, G. and {Fiore}, F.},
        title = "{The WISSH quasars project. III. X-ray properties of hyper-luminous quasars}",
      journal = {\aap},
         year = 2017,
        month = dec,
       volume = {608},
          eid = {A51},
        pages = {A51},
          doi = {10.1051/0004-6361/201731314},
archivePrefix = {arXiv},
       eprint = {1708.00452},
 primaryClass = {astro-ph.GA},
       adsurl = {https://ui.adsabs.harvard.edu/abs/2017A&A...608A..51M}
}

@ARTICLE{2021ApJ...914L..14R,
       author = {{Richards}, Gordon T. and {Plotkin}, Richard M. and {Hewett}, Paul C. and {Rankine}, Amy L. and {Rivera}, Angelica B. and {Shen}, Yue and {Shemmer}, Ohad},
        title = "{A Novel Test of Quasar Orientation}",
      journal = {\apjl},
         year = 2021,
        month = jun,
       volume = {914},
       number = {1},
          eid = {L14},
        pages = {L14},
          doi = {10.3847/2041-8213/ac0256},
archivePrefix = {arXiv},
       eprint = {2106.02633},
 primaryClass = {astro-ph.GA},
       adsurl = {https://ui.adsabs.harvard.edu/abs/2021ApJ...914L..14R}
}

@ARTICLE{2005AJ....129..567G,
       author = {{Gallagher}, S.~C. and {Richards}, Gordon T. and {Hall}, Patrick B. and {Brandt}, W.~N. and {Schneider}, Donald P. and {Vanden Berk}, Daniel E.},
        title = "{X-Ray Insights into Interpreting C IV Blueshifts and Optical/Ultraviolet Continua}",
      journal = {\aj},
         year = 2005,
        month = feb,
       volume = {129},
       number = {2},
        pages = {567-577},
          doi = {10.1086/426913},
archivePrefix = {arXiv},
       eprint = {astro-ph/0410641},
 primaryClass = {astro-ph},
       adsurl = {https://ui.adsabs.harvard.edu/abs/2005AJ....129..567G}
}

@ARTICLE{2011AJ....142..130K,
       author = {{Kruczek}, Nicholas E. and {Richards}, Gordon T. and {Gallagher}, S.~C. and {Deo}, Rajesh P. and {Hall}, Patrick B. and {Hewett}, Paul C. and {Leighly}, Karen M. and {Krawczyk}, Coleman M. and {Proga}, Daniel},
        title = "{C IV Emission and the Ultraviolet through X-Ray Spectral Energy Distribution of Radio-quiet Quasars}",
      journal = {\aj},
         year = 2011,
        month = oct,
       volume = {142},
       number = {4},
          eid = {130},
        pages = {130},
          doi = {10.1088/0004-6256/142/4/130},
archivePrefix = {arXiv},
       eprint = {1109.1515},
 primaryClass = {astro-ph.CO},
       adsurl = {https://ui.adsabs.harvard.edu/abs/2011AJ....142..130K}
}

@ARTICLE{2022ApJ...931..154R,
       author = {{Rivera}, Angelica B. and {Richards}, Gordon T. and {Gallagher}, Sarah C. and {McCaffrey}, Trevor V. and {Rankine}, Amy L. and {Hewett}, Paul C. and {Shemmer}, Ohad},
        title = "{Exploring Changes in Quasar Spectral Energy Distributions across C IV Parameter Space}",
      journal = {\apj},
         year = 2022,
        month = jun,
       volume = {931},
       number = {2},
          eid = {154},
        pages = {154},
          doi = {10.3847/1538-4357/ac6a5d},
       adsurl = {https://ui.adsabs.harvard.edu/abs/2022ApJ...931..154R}
}

@ARTICLE{2004ApJ...611..125L,
       author = {{Leighly}, Karen M.},
        title = "{Hubble Space Telescope STIS Ultraviolet Spectral Evidence of Outflow in Extreme Narrow-Line Seyfert 1 Galaxies. II. Modeling and Interpretation}",
      journal = {\apj},
         year = 2004,
        month = aug,
       volume = {611},
       number = {1},
        pages = {125-152},
          doi = {10.1086/422089},
archivePrefix = {arXiv},
       eprint = {astro-ph/0402452},
 primaryClass = {astro-ph},
       adsurl = {https://ui.adsabs.harvard.edu/abs/2004ApJ...611..125L}
}

@ARTICLE{1991ApJ...374..344K,
       author = {{Kraft}, Ralph P. and {Burrows}, David N. and {Nousek}, John A.},
        title = "{Determination of Confidence Limits for Experiments with Low Numbers of Counts}",
      journal = {\apj},
         year = 1991,
        month = jun,
       volume = {374},
        pages = {344},
          doi = {10.1086/170124},
       adsurl = {https://ui.adsabs.harvard.edu/abs/1991ApJ...374..344K}
}

@ARTICLE{2010ApJ...724..762W,
       author = {{Wu}, Jianfeng and {Brandt}, W.~N. and {Comins}, M.~L. and {Gibson}, Robert R. and {Shemmer}, Ohad and {Garmire}, Gordon P. and {Schneider}, Donald P.},
        title = "{The X-ray Properties of the Optically Brightest Mini-Bal Quasars from the Sloan Digital Sky Survey}",
      journal = {\apj},
         year = 2010,
        month = nov,
       volume = {724},
       number = {1},
        pages = {762-778},
          doi = {10.1088/0004-637X/724/1/762},
archivePrefix = {arXiv},
       eprint = {1009.3928},
 primaryClass = {astro-ph.CO},
       adsurl = {https://ui.adsabs.harvard.edu/abs/2010ApJ...724..762W}
}

@ARTICLE{2025MNRAS.542.2105H,
       author = {{Hiremath}, Pranavi and {Rankine}, Amy L. and {Aird}, James and {Brandt}, W.~N. and {Rodr{\'\i}guez Hidalgo}, Paola and {Anderson}, Scott F. and {Aydar}, Catarina and {Ricci}, Claudio and {Schneider}, Donald P. and {Vivek}, M. and {Igo}, Zsofi and {Morrison}, Sean and {Salvato}, Mara},
        title = "{X-ray selected broad absorption line quasars in SDSS-V: BALs and non-BALs span the same range of X-ray properties}",
      journal = {\mnras},
         year = 2025,
        month = sep,
       volume = {542},
       number = {3},
        pages = {2105-2127},
          doi = {10.1093/mnras/staf1352},
archivePrefix = {arXiv},
       eprint = {2508.13682},
 primaryClass = {astro-ph.GA},
       adsurl = {https://ui.adsabs.harvard.edu/abs/2025MNRAS.542.2105H}
}

@ARTICLE{2009A&A...495..421B,
       author = {{Bianchi}, S. and {Guainazzi}, M. and {Matt}, G. and {Fonseca Bonilla}, N. and {Ponti}, G.},
        title = "{CAIXA: a catalogue of AGN in the XMM-Newton archive. I. Spectral analysis}",
      journal = {\aap},
         year = 2009,
        month = feb,
       volume = {495},
       number = {2},
        pages = {421-430},
          doi = {10.1051/0004-6361:200810620},
archivePrefix = {arXiv},
       eprint = {0811.1126},
 primaryClass = {astro-ph},
       adsurl = {https://ui.adsabs.harvard.edu/abs/2009A&A...495..421B}
}

@ARTICLE{2009ApJS..183...17Y,
       author = {{Young}, M. and {Elvis}, M. and {Risaliti}, G.},
        title = "{The Fifth Data Release Sloan Digital Sky Survey/XMM-Newton Quasar Survey}",
      journal = {\apjs},
         year = 2009,
        month = jul,
       volume = {183},
       number = {1},
        pages = {17-32},
          doi = {10.1088/0067-0049/183/1/17},
archivePrefix = {arXiv},
       eprint = {0905.0496},
 primaryClass = {astro-ph.HE},
       adsurl = {https://ui.adsabs.harvard.edu/abs/2009ApJS..183...17Y}
}

@ARTICLE{2018A&A...617A..81V,
       author = {{Vietri}, G. and {Piconcelli}, E. and {Bischetti}, M. and {Duras}, F. and {Martocchia}, S. and {Bongiorno}, A. and {Marconi}, A. and {Zappacosta}, L. and {Bisogni}, S. and {Bruni}, G. and {Brusa}, M. and {Comastri}, A. and {Cresci}, G. and {Feruglio}, C. and {Giallongo}, E. and {La Franca}, F. and {Mainieri}, V. and {Mannucci}, F. and {Ricci}, F. and {Sani}, E. and {Testa}, V. and {Tombesi}, F. and {Vignali}, C. and {Fiore}, F.},
        title = "{The WISSH quasars project. IV. Broad line region versus kiloparsec-scale winds}",
      journal = {\aap},
         year = 2018,
        month = sep,
       volume = {617},
          eid = {A81},
        pages = {A81},
          doi = {10.1051/0004-6361/201732335},
archivePrefix = {arXiv},
       eprint = {1802.03423},
 primaryClass = {astro-ph.GA},
       adsurl = {https://ui.adsabs.harvard.edu/abs/2018A&A...617A..81V}
}

@ARTICLE{2019A&A...630A..94G,
       author = {{Giustini}, Margherita and {Proga}, Daniel},
        title = "{A global view of the inner accretion and ejection flow around super massive black holes. Radiation-driven accretion disk winds in a physical context}",
      journal = {\aap},
         year = 2019,
        month = oct,
       volume = {630},
          eid = {A94},
        pages = {A94},
          doi = {10.1051/0004-6361/201833810},
archivePrefix = {arXiv},
       eprint = {1904.07341},
 primaryClass = {astro-ph.GA},
       adsurl = {https://ui.adsabs.harvard.edu/abs/2019A&A...630A..94G}
}

@article{10.1093/biomet/26.4.404,
    author = {Clopper, C. J. and Pearson, E. S.},
    title = {THE USE OF CONFIDENCE OR FIDUCIAL LIMITS ILLUSTRATED IN THE CASE OF THE BINOMIAL},
    journal = {Biometrika},
    volume = {26},
    number = {4},
    pages = {404-413},
    year = {1934},
    month = {12},
    issn = {0006-3444},
    doi = {10.1093/biomet/26.4.404},
    url = {https://doi.org/10.1093/biomet/26.4.404},
    eprint = {https://academic.oup.com/biomet/article-pdf/26/4/404/823407/26-4-404.pdf},
}

@article{10.1214/ss/1009213286,
author = {Lawrence D. Brown and T. Tony Cai and Anirban DasGupta},
title = {{Interval Estimation for a Binomial Proportion}},
volume = {16},
journal = {Stat. Sci.},
number = {2},
publisher = {Institute of Mathematical Statistics},
pages = {101 -- 133},
year = {2001},
doi = {10.1214/ss/1009213286},
URL = {https://doi.org/10.1214/ss/1009213286}
}

@ARTICLE{2021MNRAS.504.5556T,
       author = {{Timlin, J. D., III} and {Brandt}, W.~N. and {Laor}, Ari},
        title = "{What controls the UV-to-X-ray continuum shape in quasars?}",
      journal = {\mnras},
         year = 2021,
        month = jul,
       volume = {504},
       number = {4},
        pages = {5556-5574},
          doi = {10.1093/mnras/stab1217},
archivePrefix = {arXiv},
       eprint = {2104.13938},
 primaryClass = {astro-ph.HE},
       adsurl = {https://ui.adsabs.harvard.edu/abs/2021MNRAS.504.5556T}
}

@ARTICLE{2023MNRAS.524.5497S,
       author = {{Stepney}, Matthew and {Banerji}, Manda and {Hewett}, Paul C. and {Temple}, Matthew J. and {Rankine}, Amy L. and {Matthews}, James H. and {Richards}, Gordon T.},
        title = "{No redshift evolution in the rest-frame ultraviolet emission line properties of quasars from z = 1.5 to z = 4.0}",
      journal = {\mnras},
         year = 2023,
        month = oct,
       volume = {524},
       number = {4},
        pages = {5497-5513},
          doi = {10.1093/mnras/stad2060},
archivePrefix = {arXiv},
       eprint = {2307.02962},
 primaryClass = {astro-ph.GA},
       adsurl = {https://ui.adsabs.harvard.edu/abs/2023MNRAS.524.5497S}
}

@ARTICLE{2010MNRAS.405.2302H,
       author = {{Hewett}, Paul C. and {Wild}, Vivienne},
        title = "{Improved redshifts for SDSS quasar spectra}",
      journal = {\mnras},
         year = 2010,
        month = jul,
       volume = {405},
       number = {4},
        pages = {2302-2316},
          doi = {10.1111/j.1365-2966.2010.16648.x},
archivePrefix = {arXiv},
       eprint = {1003.3017},
 primaryClass = {astro-ph.CO},
       adsurl = {https://ui.adsabs.harvard.edu/abs/2010MNRAS.405.2302H}
}

\begin{appendix}
\section{Gallery of optical spectra} \label{optspectrall}
Figure~\ref{fig:optspectrall} presents the modelling of the C\,{\sc iv} line of the SDSS spectra with multiple components, as discussed in Section~\ref{sample}. The spectral fits were performed through a custom-made code, based on the IDL MPFIT package \citepads{2009ASPC..411..251M}, which takes advantage of the Levenberg–Marquardt technique \citepads{1978LNM...630..105M} to solve the least-squares problem. We modelled the various emission lines present in the 1350\,--1700\,\AA\ range, chiefly the O\,{\sc iv}]+Si\,{\sc iv}$~\lambda$1400\,\AA\ blend (Si\,{\sc iv} in short) and the C\,{\sc iv}, using Gaussian or Lorentzian profiles. In case an outflow component was detected in both Si\,{\sc iv} and C\,{\sc iv}, these two components were not tied together. In a handful of cases where the spectra showed the presence of the Fe\,{\sc ii} pseudo-continuum, we modelled it using some purposefully made \textsc{CLOUDY} \citepads{2013RMxAA..49..137F} templates (see Section~3.1 in \citeads{2023A&A...677A.111T} for more details). 

Our sample selection is designed to exclude quasars with strong broad absorption features, where such characteristics would impede a reliable continuum fit. We note that two sources in our main dataset, J090924.01+000211.0 and J093514.71+033545.7, are flagged as broad absorption line (BAL) quasars in the \citetads{2022ApJS..263...42W} catalogue. However, our final selection included a visual inspection, which confirmed that both objects exhibit relatively blue continua without the deep, wide absorption troughs that typically challenge spectral fitting and the determination of the continuum. Therefore, given that our continuum determination is robust and there is no established correlation between these specific types of `blue' BAL quasars and X-ray absorption (see, e.g. \citeads{2010ApJ...724..762W}; \citeads{2019A&A...632A.109N}; \citeads{2025MNRAS.542.2105H}), we retained them in the sample. Separately, we also identified absorption features in the UV spectra of J111800.50+195853.4 and J103005.10+132531.1. These objects are not classified as BAL quasars in the \citetads{2022ApJS..263...42W} catalogue. The absorption systems identified in our spectral fitting for these two sources are consistent with narrower associated absorbers, which are distinct from the BAL population. Furthermore, the spectral fitting procedure takes into account these absorption features. Thus, for these two cases as well, the continuum determination is reliable. We are therefore confident that the inclusion of those four objects does not bias our results.

We report the fitted lines as labels in the corresponding figures. There, we colour-coded several components employed in the fit, as shown in the legend. The reported C\,{\sc iv} outflow velocities are measured as the shift of the outflow component with respect to the BLR component. The dashed vertical lines indicate the expected position for the fitted lines according to the redshift reported in the catalogue. The shaded light grey regions are telluric bands, narrow absorption lines, or bad pixels and are therefore excluded from the fit.

   \begin{figure}[h!]
   \centering
   \includegraphics[width=\hsize]{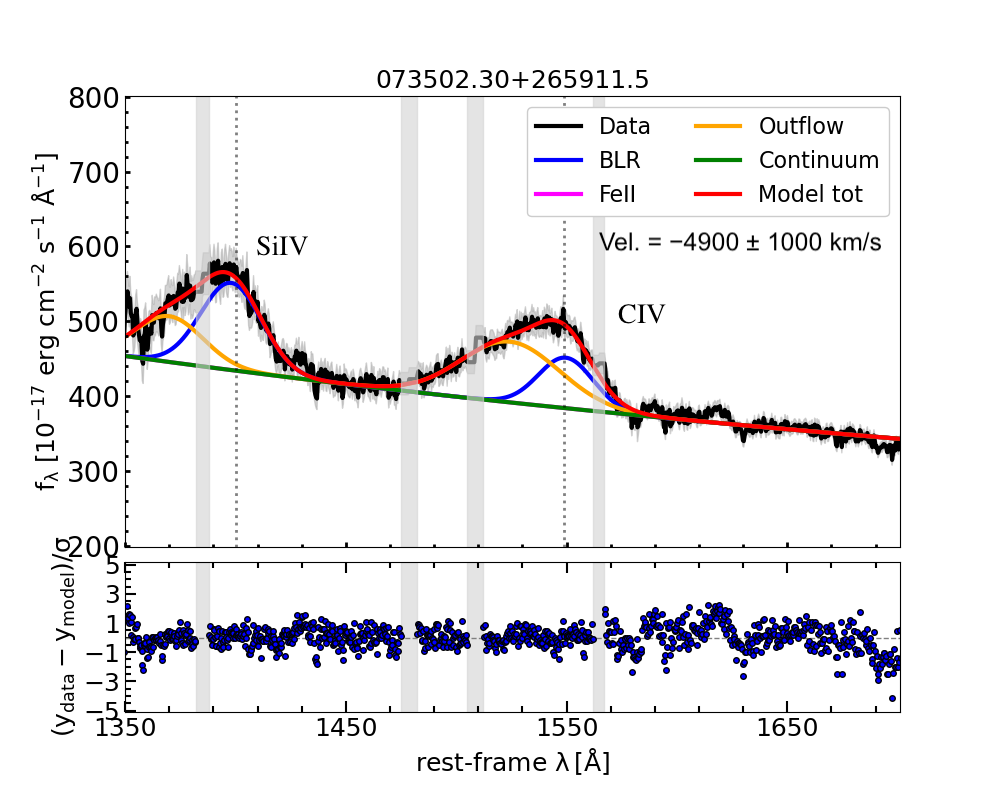}\\
   \includegraphics[width=\hsize]{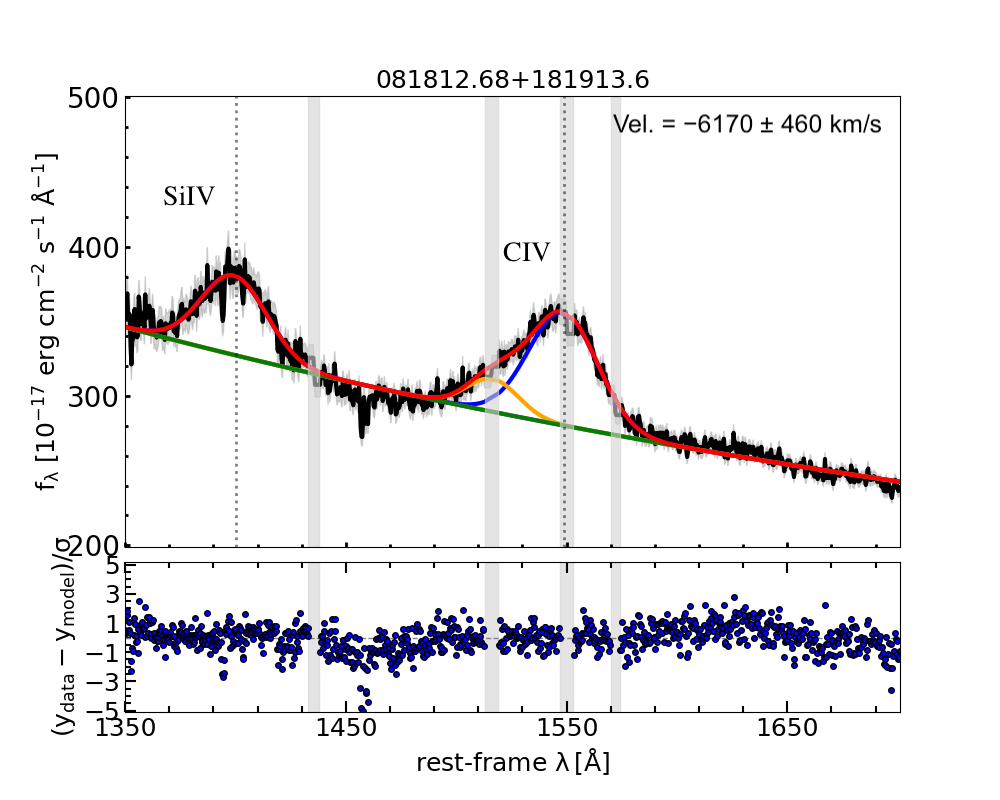}\\
   \includegraphics[width=\hsize]{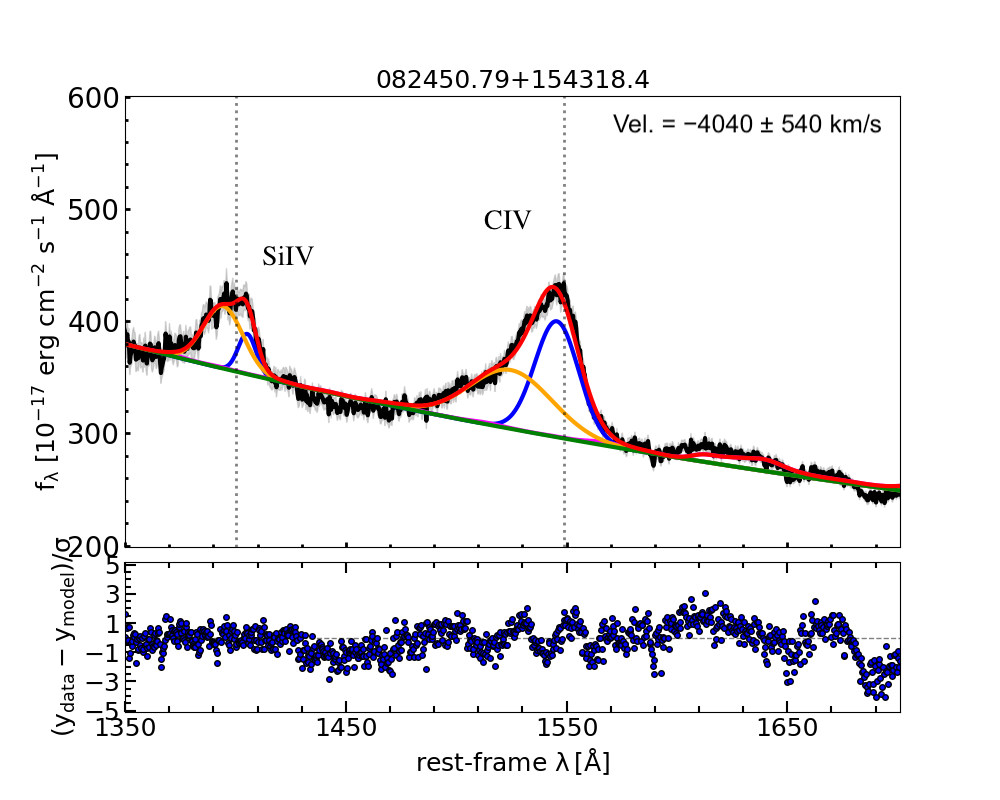}
      \caption{Analysis of C\,{\sc iv} spectral region of SDSS spectra, as discussed in Section~\ref{main_data}. Legend as in Figure~\ref{fig:optspectrum}.
              }
         \label{fig:optspectrall}
   \end{figure}

   \begin{figure*}
   \addtocounter{figure}{-1}
   \centering
   \begin{tabular}{c c}
   \includegraphics[width=0.5\hsize]{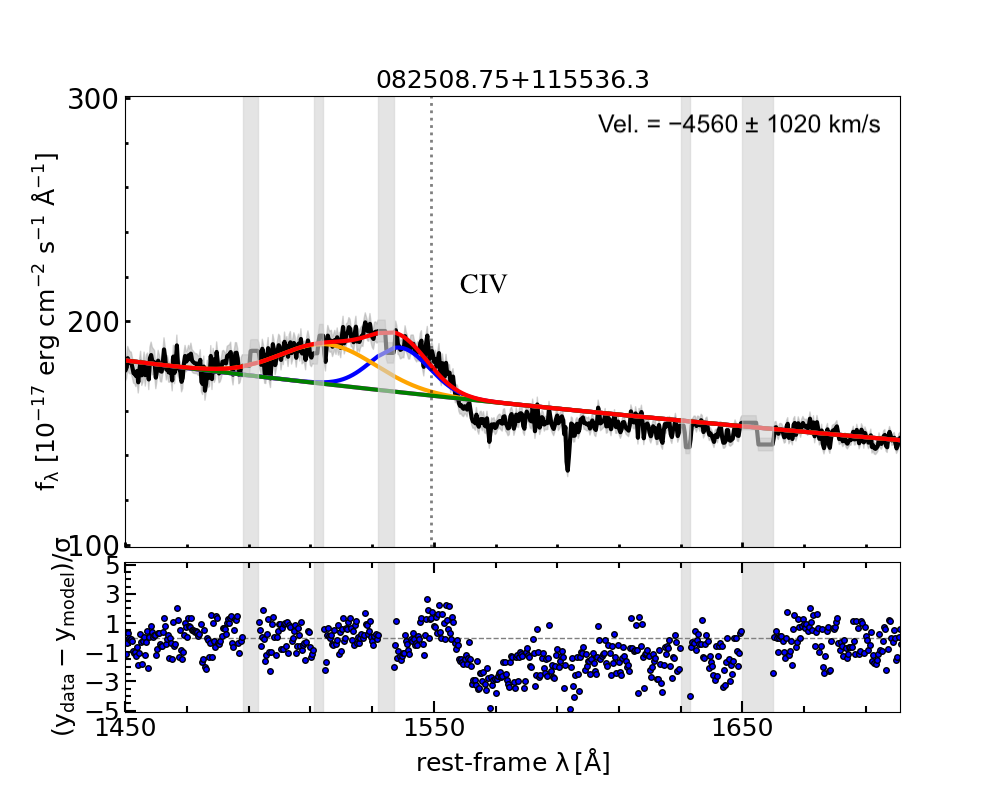} & \includegraphics[width=0.5\hsize]{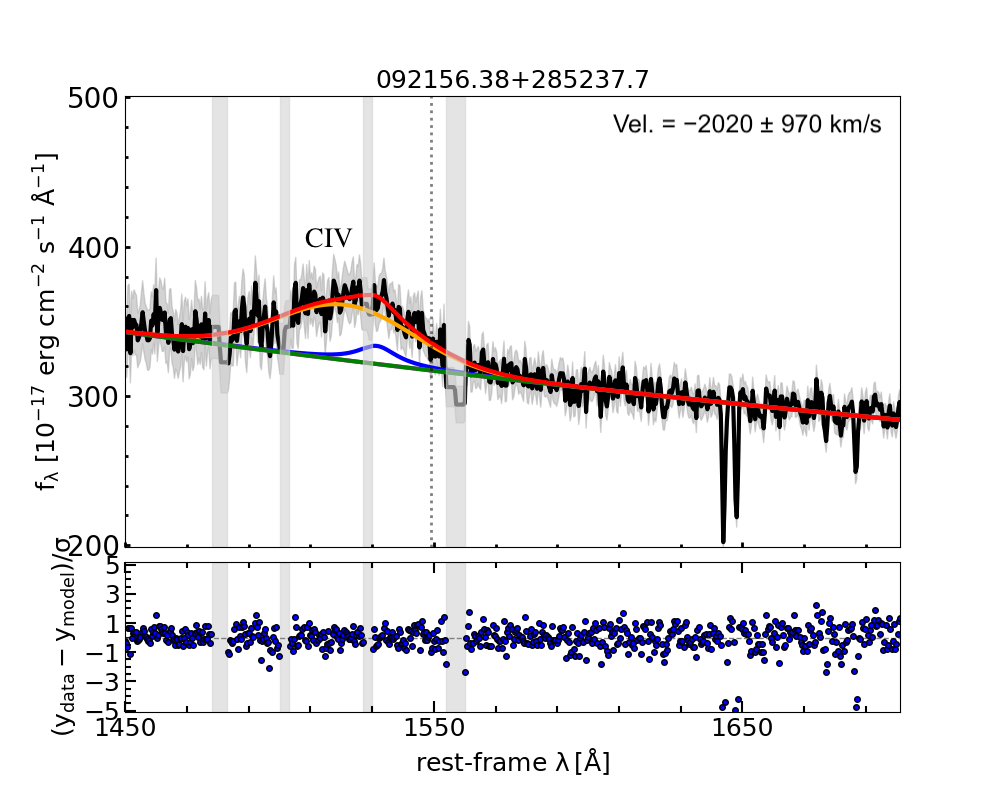}\\
   \includegraphics[width=0.5\hsize]{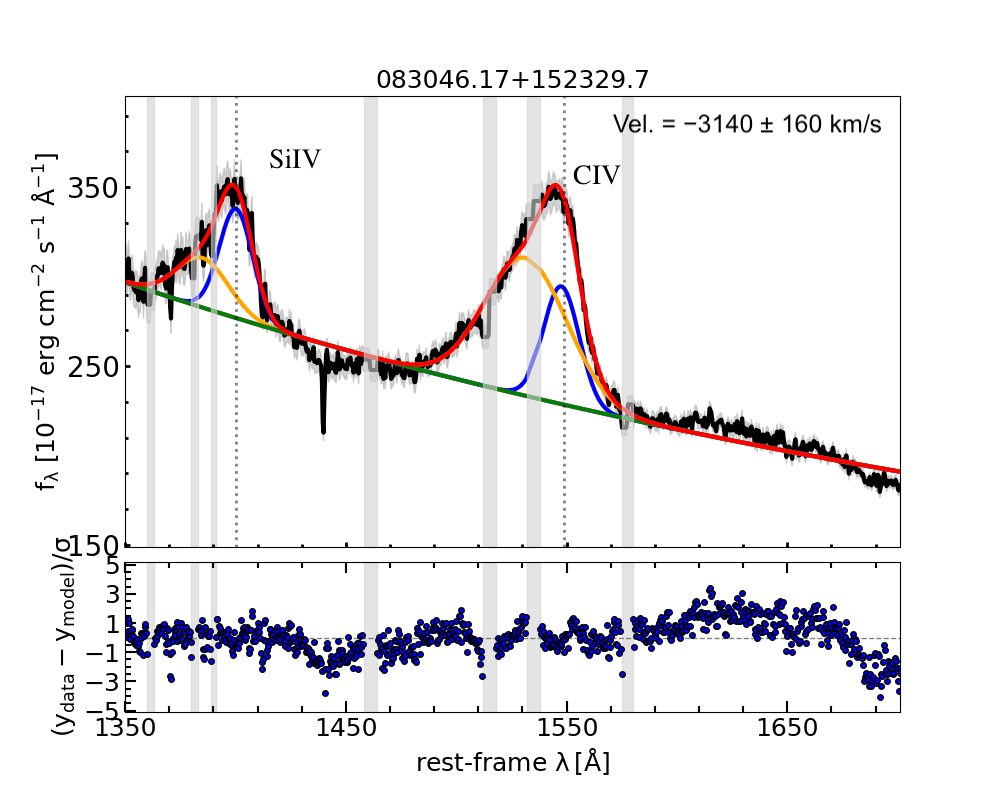} & \includegraphics[width=0.5\hsize]{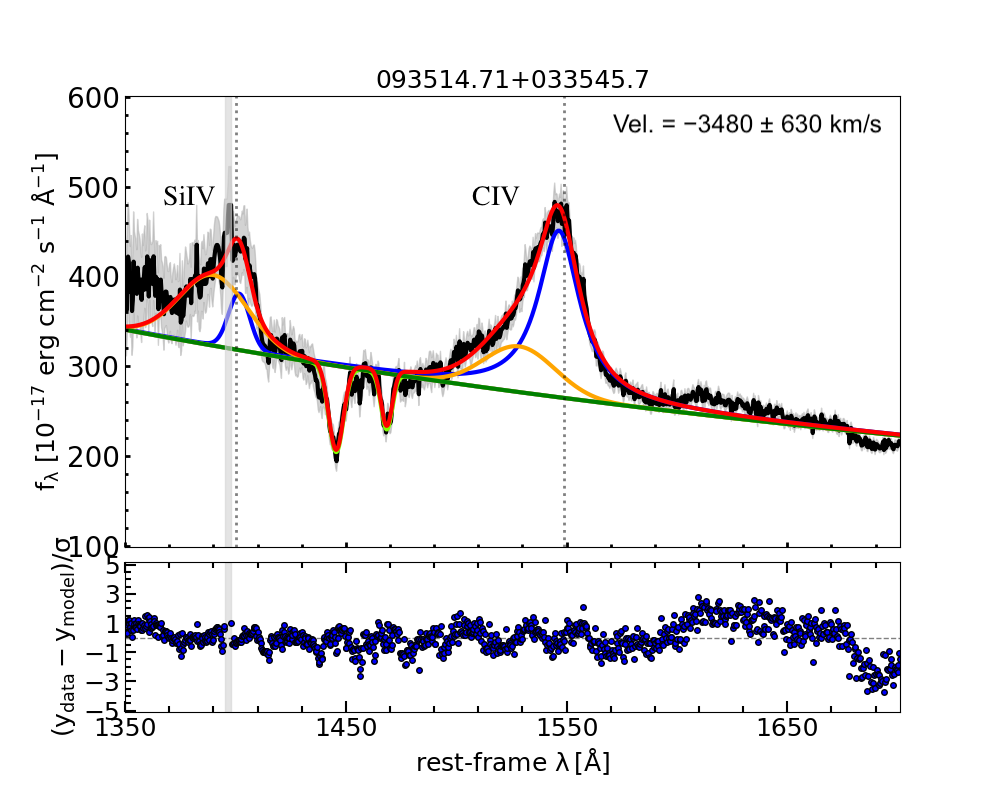}\\
   \includegraphics[width=0.5\hsize]{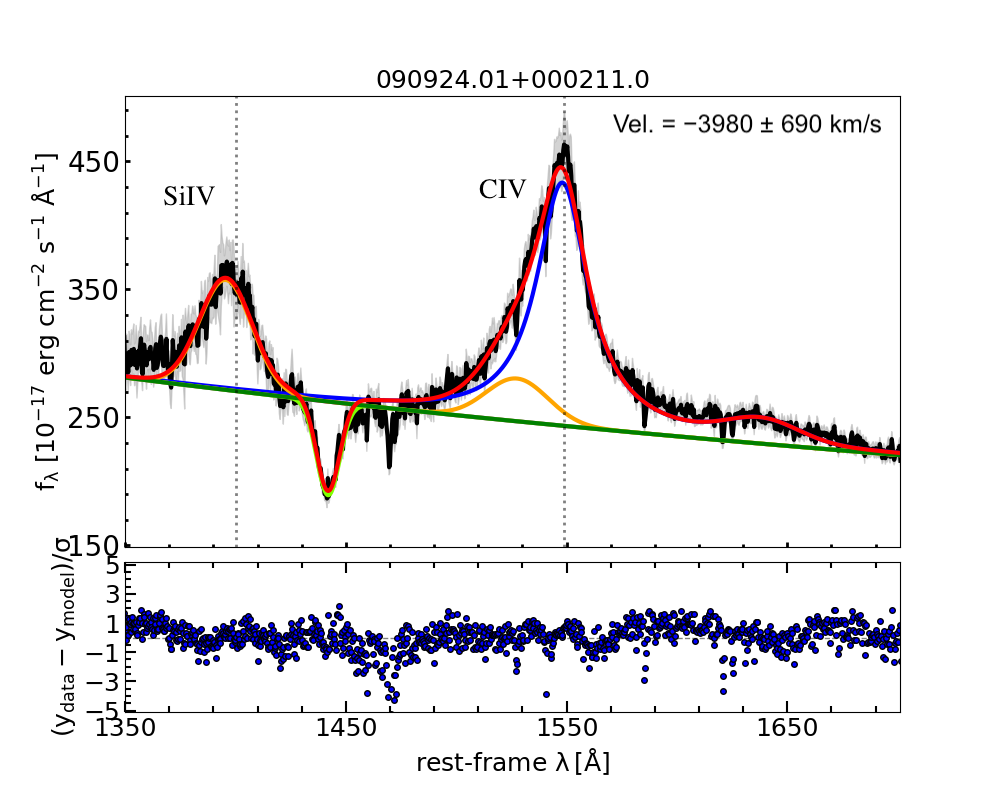} & \includegraphics[width=0.5\hsize]{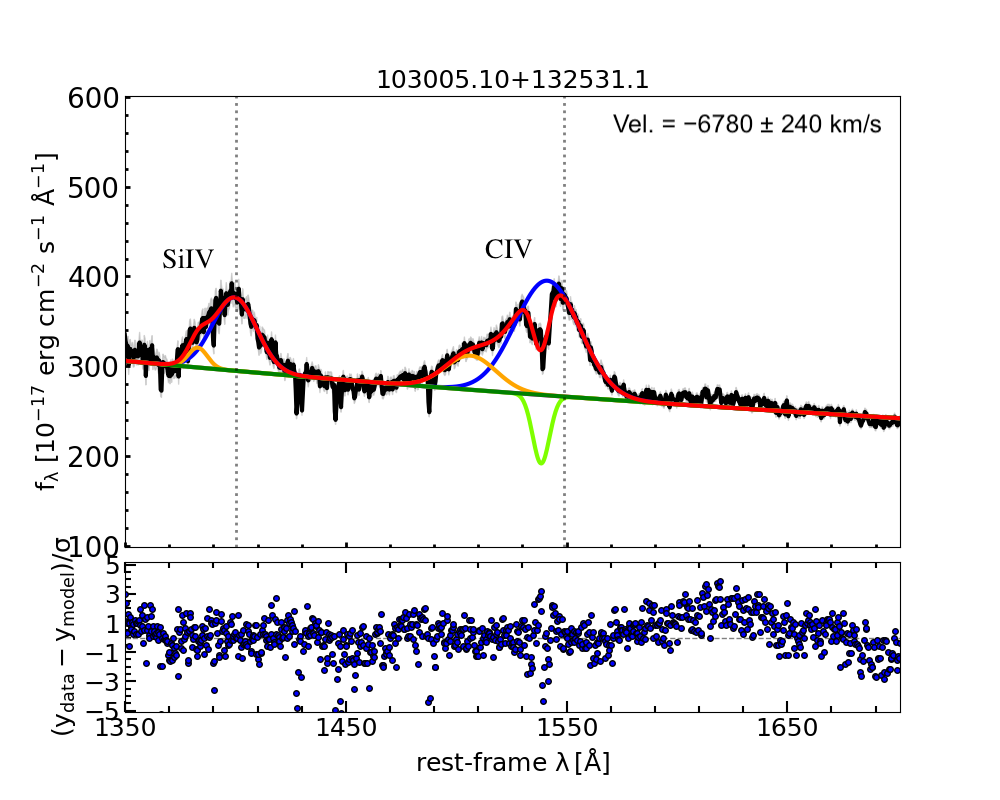}
   \end{tabular}
      \caption{continued.
              }
   \end{figure*}

   \begin{figure*}
   \addtocounter{figure}{-1}
   \centering
   \begin{tabular}{c c}
   \includegraphics[width=0.5\hsize]{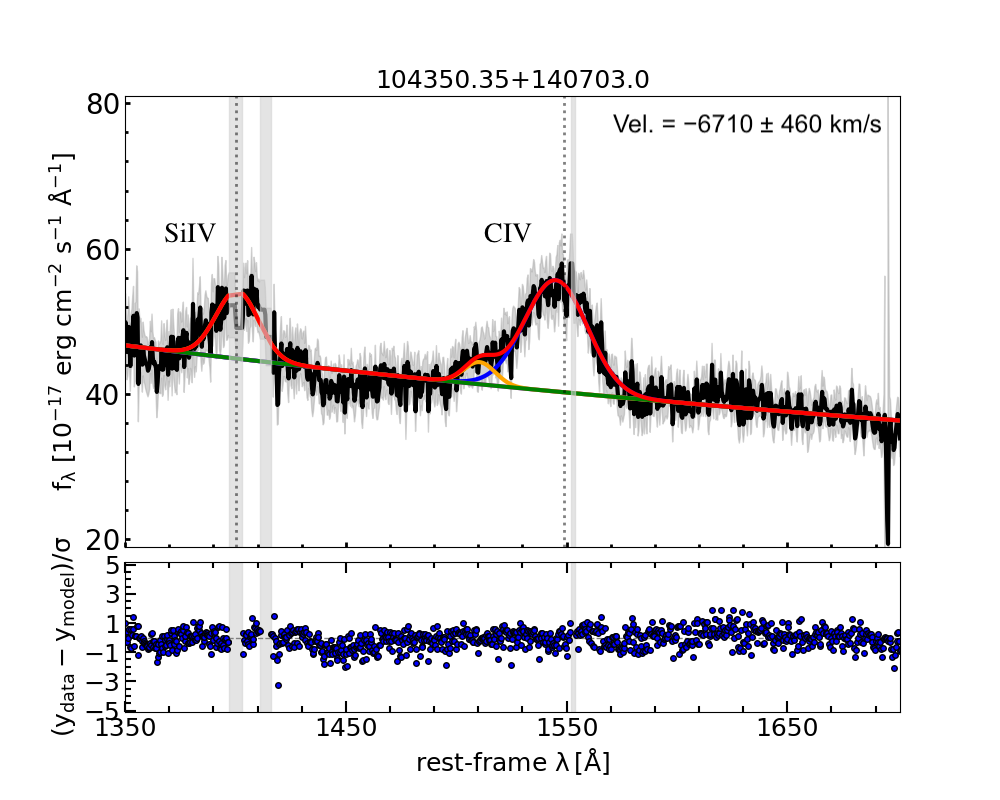} & \includegraphics[width=0.5\hsize]{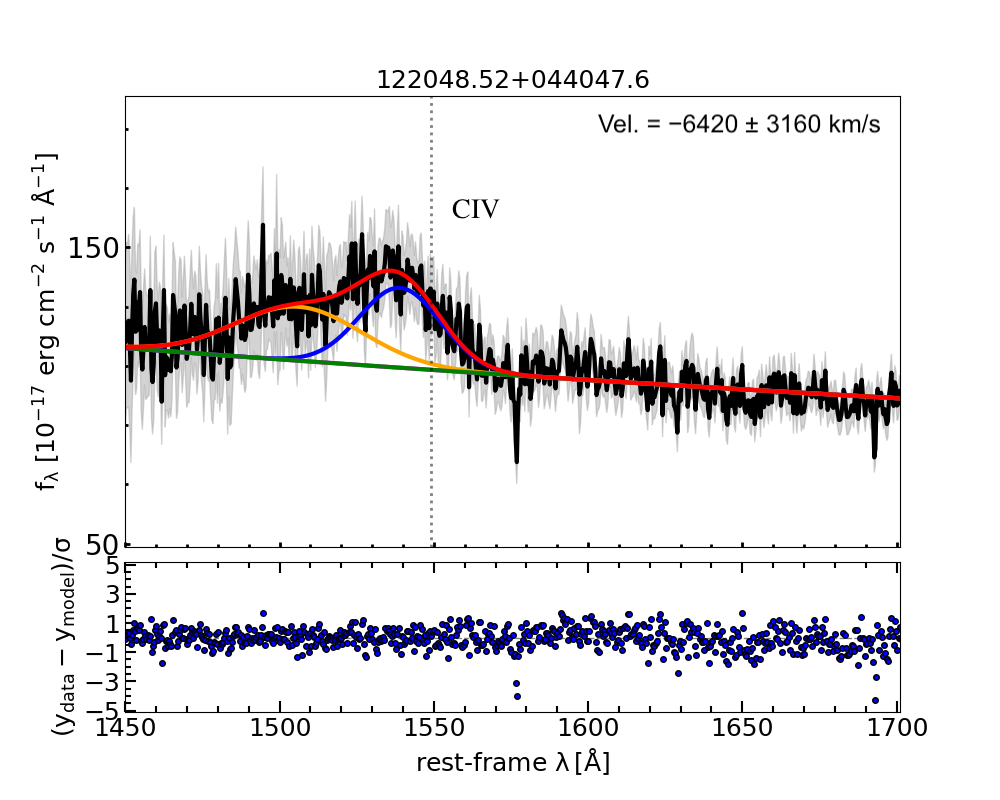}\\
   \includegraphics[width=0.5\hsize]{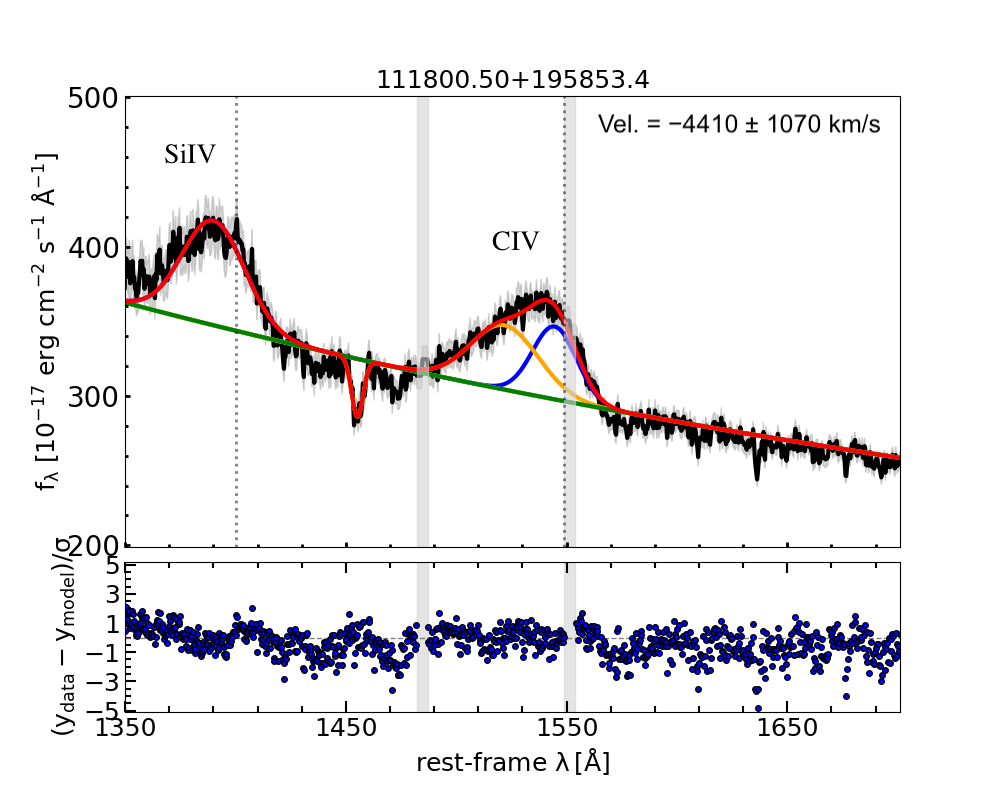} & \includegraphics[width=0.5\hsize]{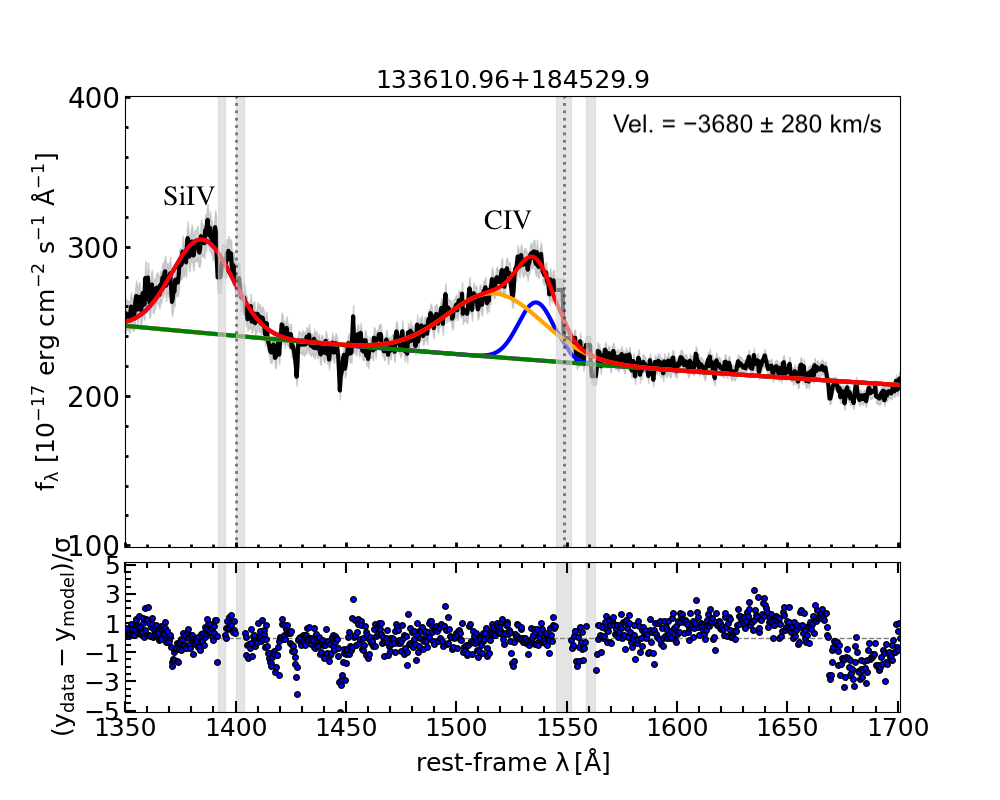}
   \end{tabular}
      \caption{continued.
              }
   \end{figure*}

\FloatBarrier
\section{Gallery of X-ray spectra}  \label{Xspectrall}
\begin{figure*}[!bh]
   \centering
   \begin{tabular}{ccc}
   \includegraphics[width=0.32\hsize]{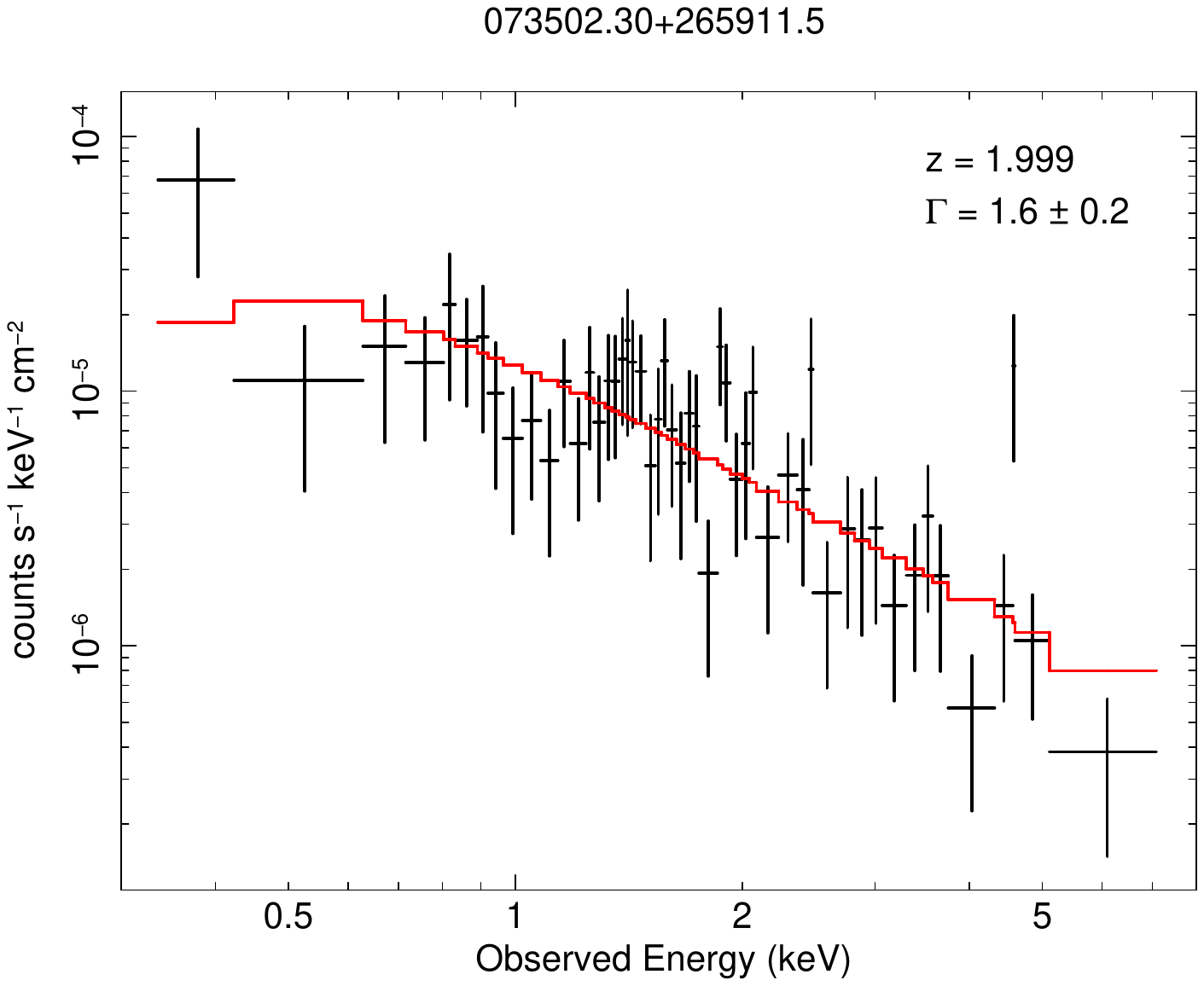} & \includegraphics[width=0.32\hsize]{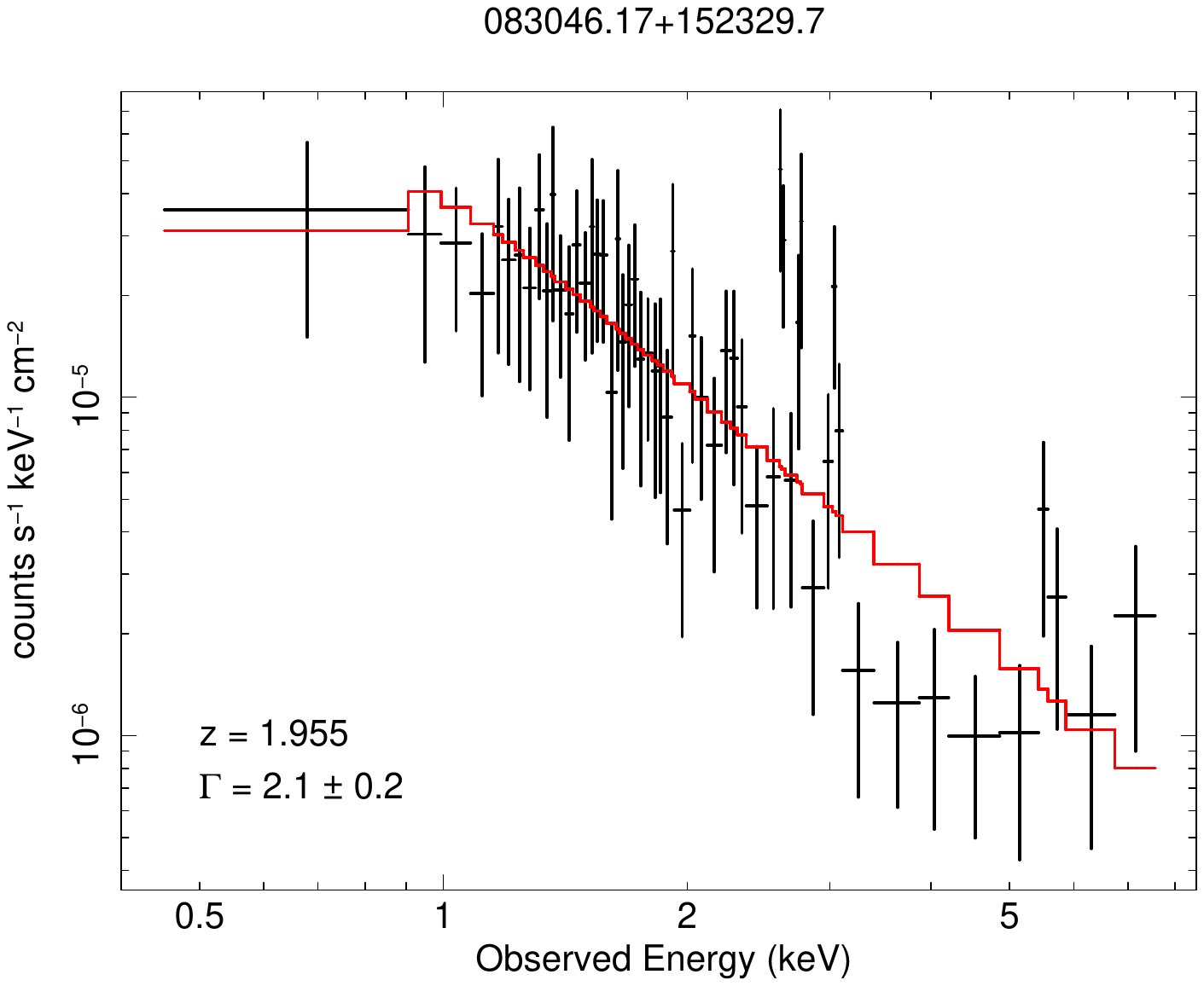} & \includegraphics[width=0.32\hsize]{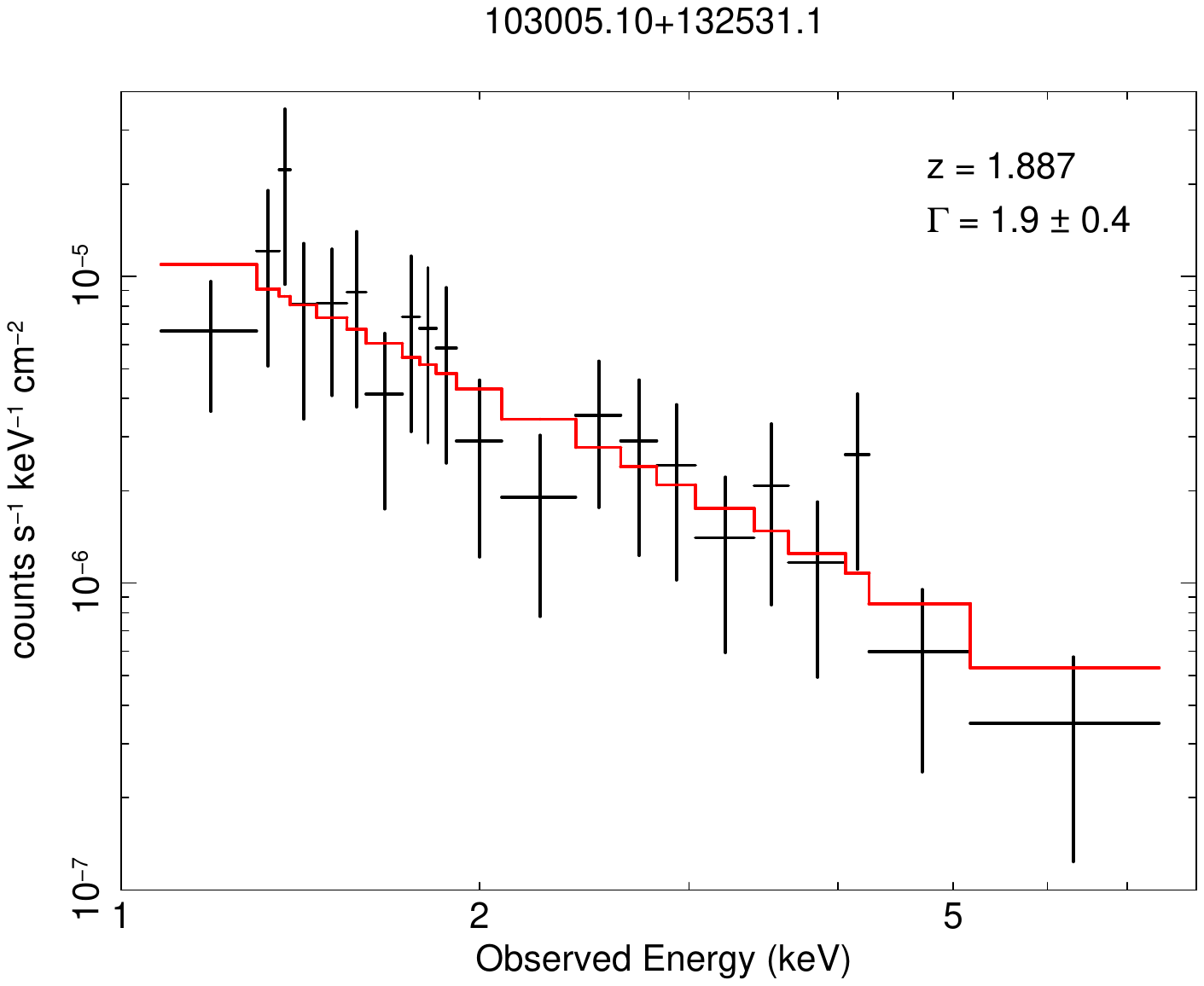}\\
   \includegraphics[width=0.32\hsize]{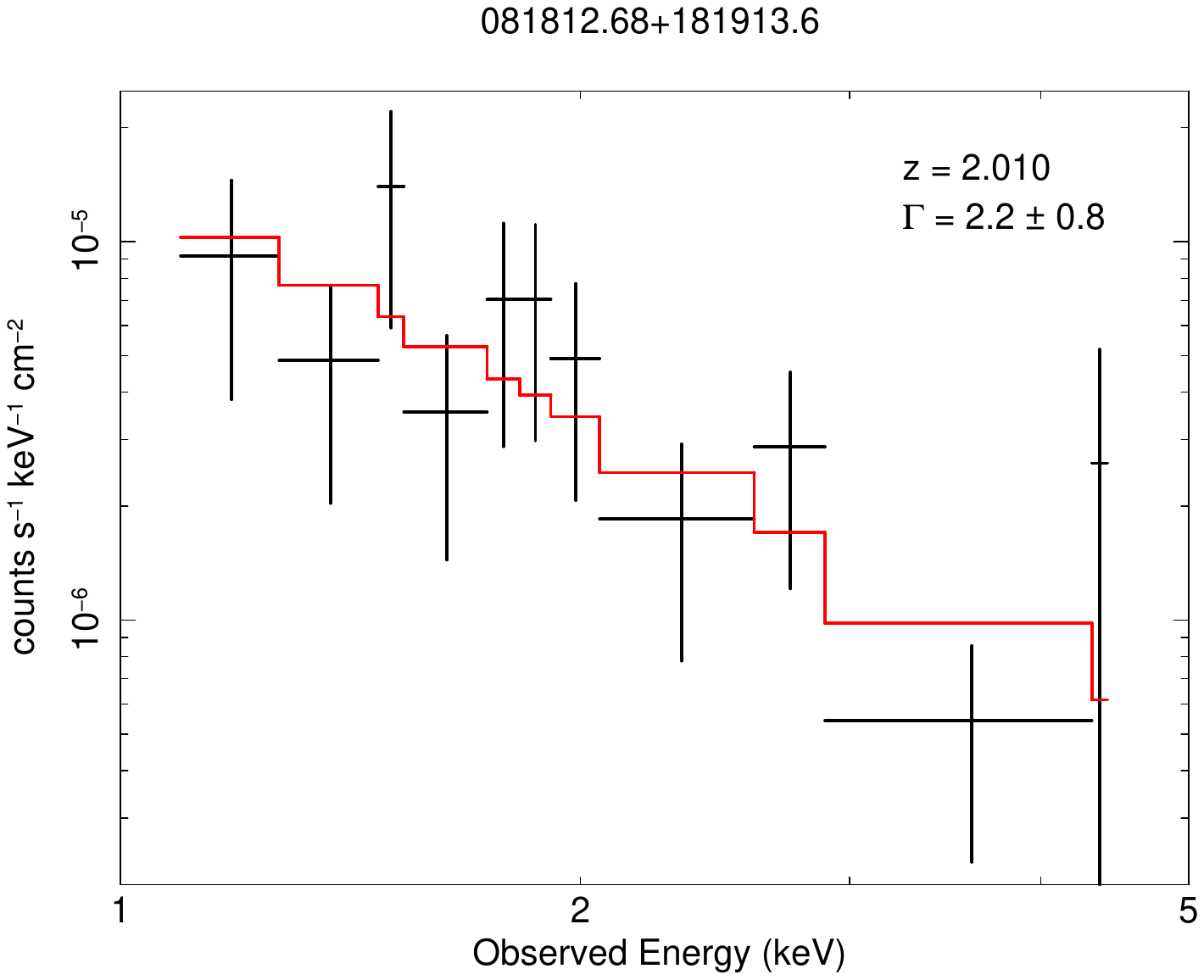} & \includegraphics[width=0.32\hsize]{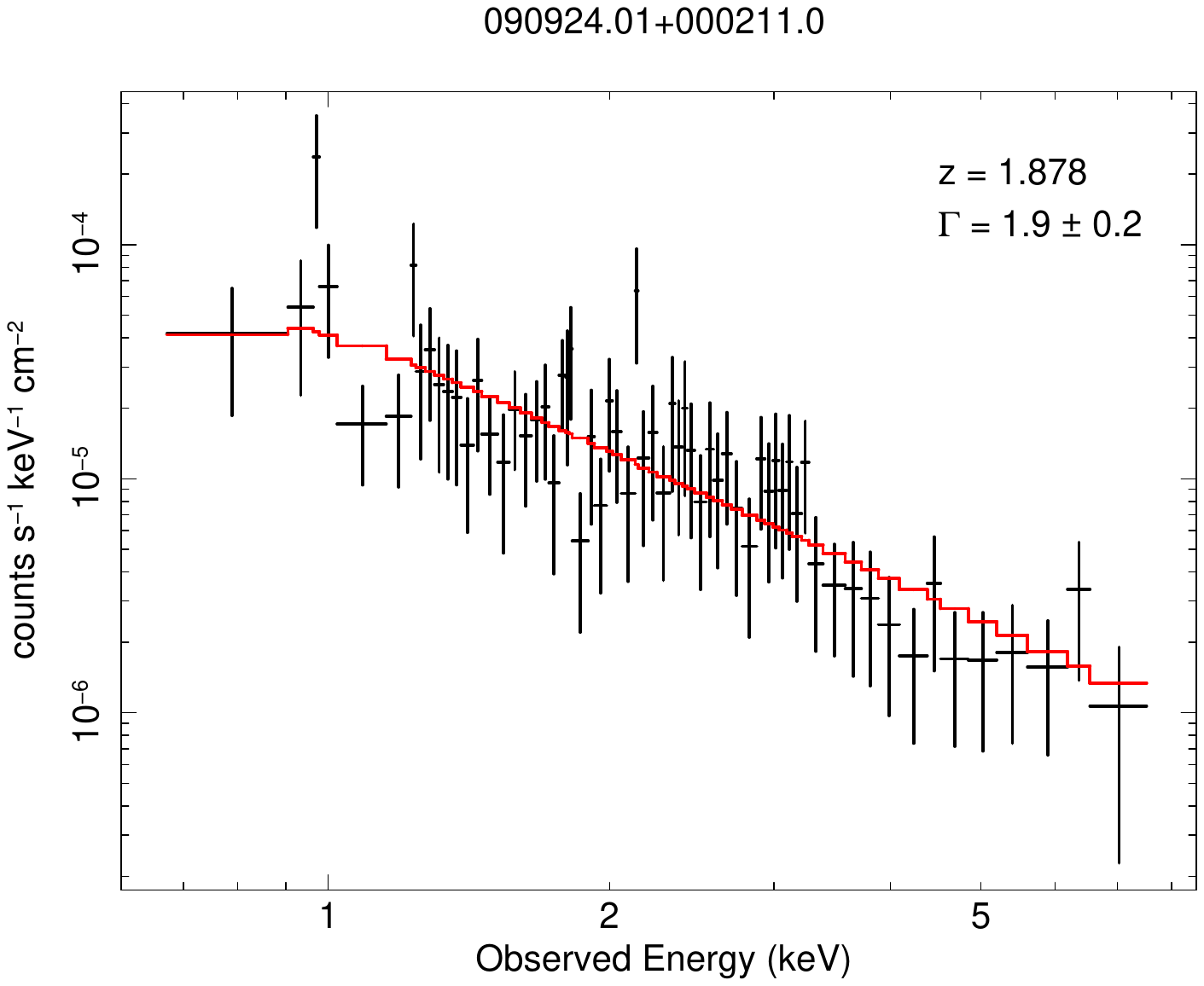} & \includegraphics[width=0.32\hsize]{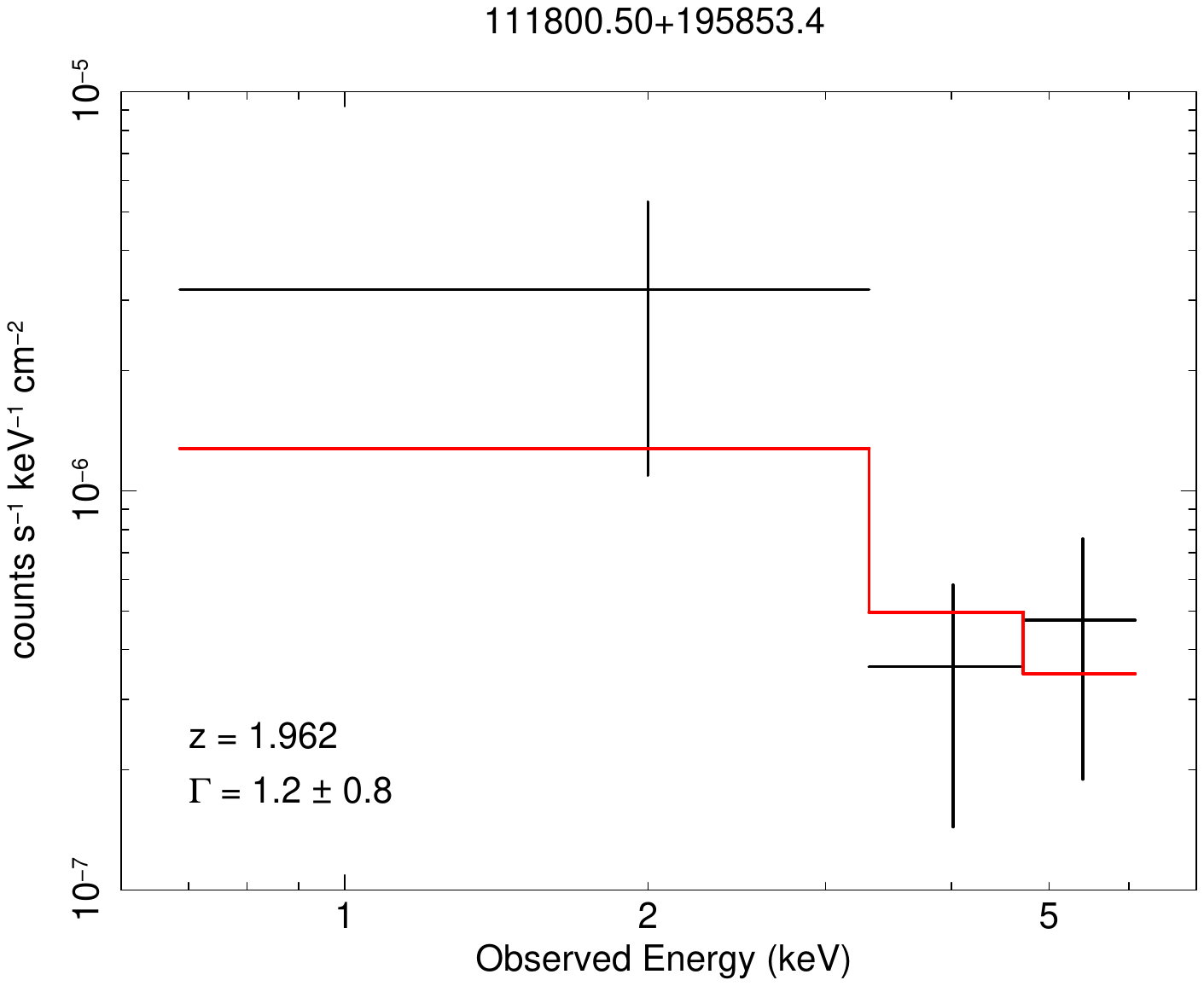}\\
   \includegraphics[width=0.32\hsize]{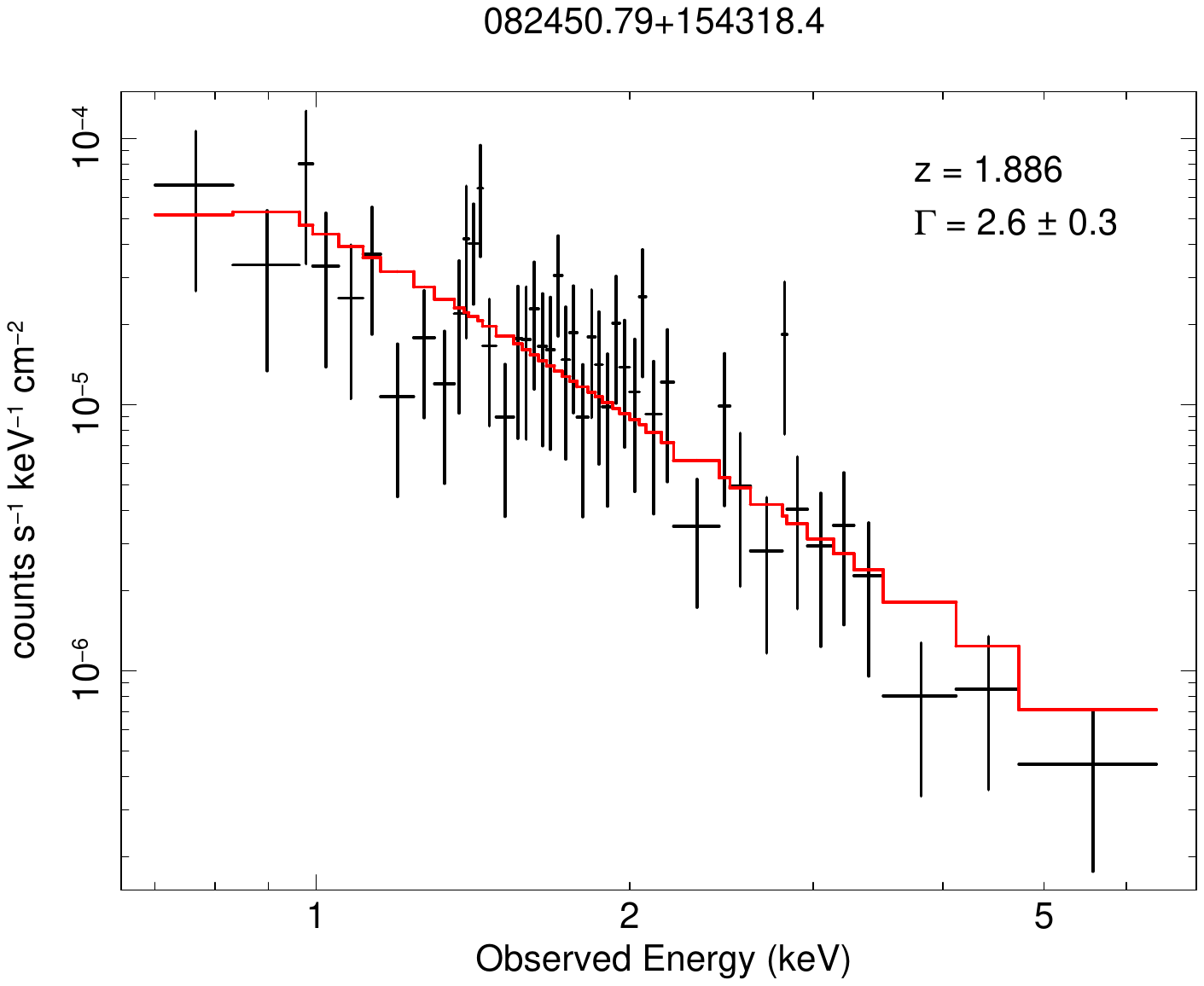} & \includegraphics[width=0.32\hsize]{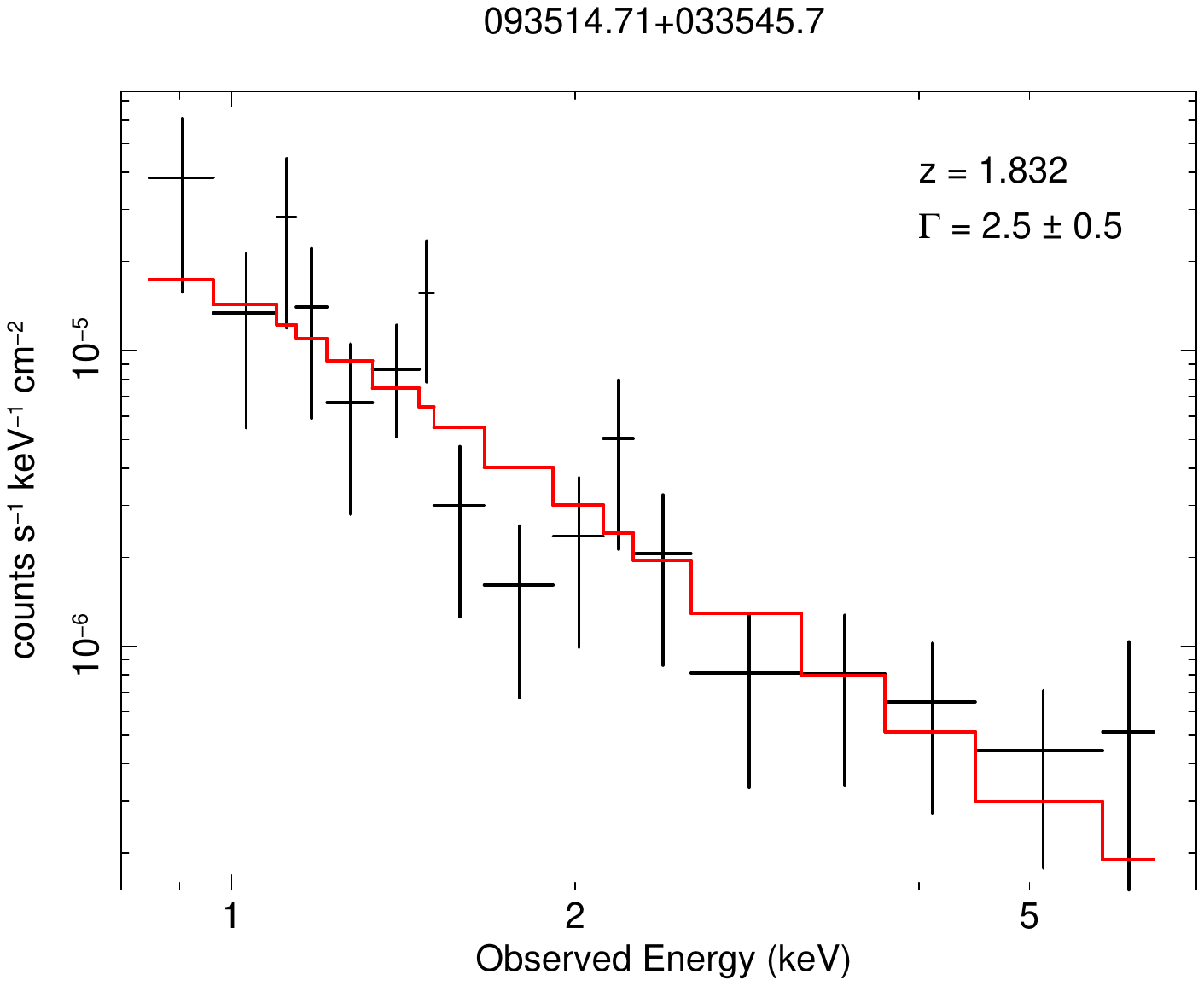} & \includegraphics[width=0.32\hsize]{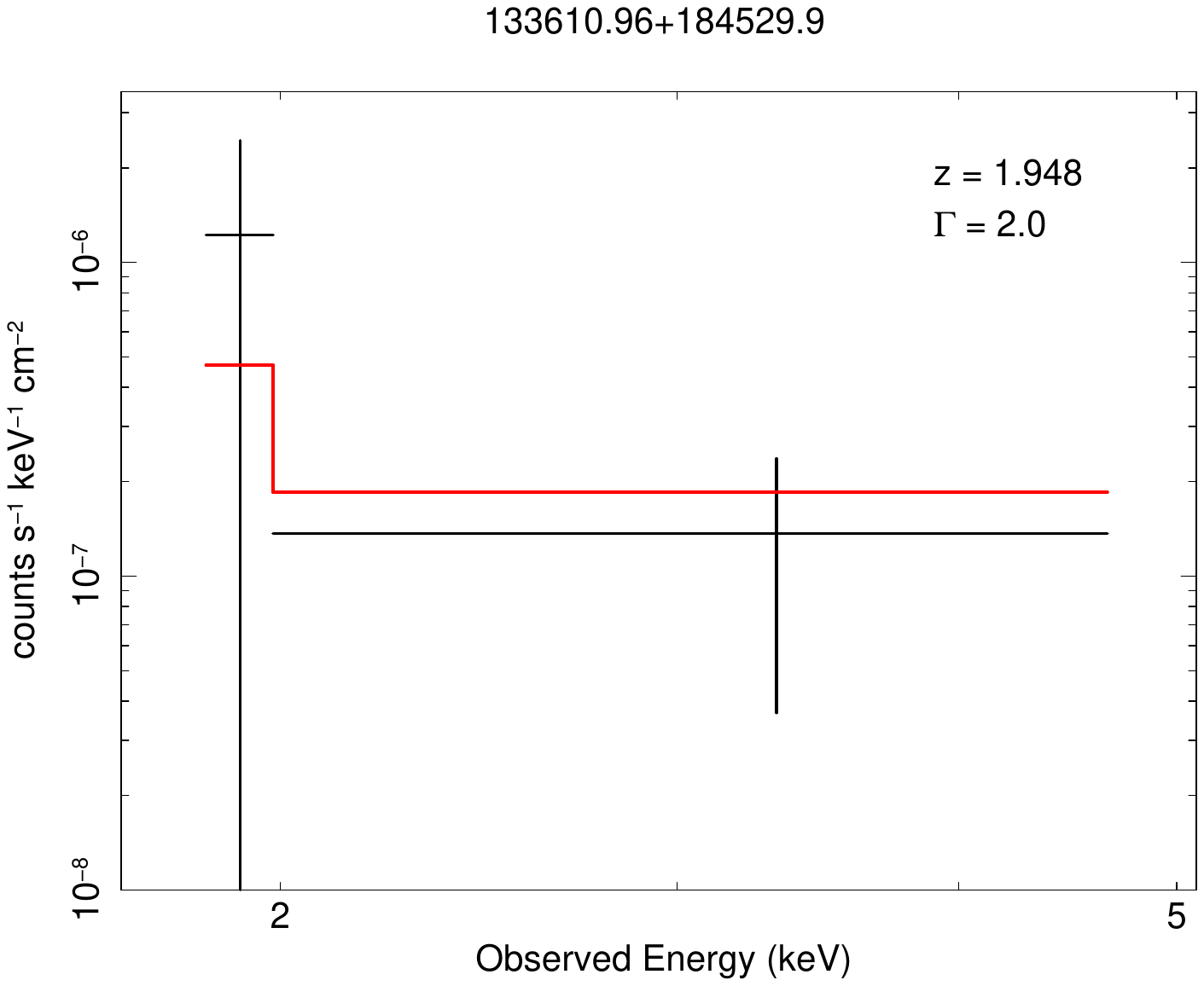}
   \end{tabular}
      \caption{Analysis of {\em Chandra} X-ray spectra, as discussed in Section~\ref{analysis}. Legend as in Figure~\ref{fig:Xspectrum}.
              }
         \label{fig:Xspectrall}
         \vspace{5cm}
   \end{figure*}

Figure~\ref{fig:Xspectrall} presents the analysis of the {\em Chandra} X-ray data for our main dataset, as discussed in Section~\ref{analysis}. To enhance visual clarity, the spectra in the plots are binned to ensure at least 1.5 counts per energy channel, except for SDSS~J133610.96+184529.9, for which the spectrum in the plot is binned to ensure at least one count per energy channel. Black crosses represent the observational data with corresponding errors, red lines represent the best-fit models.

\clearpage
\section{$L_{\mathrm{X}}$ -- $L_{\mathrm{UV}}$ relation in redshift bins} \label{relation_in_bins}
The slope of the $L_{\mathrm{X}}$ -- $L_{\mathrm{UV}}$ relation, when examined in redshift bins, does not show any clear trend with redshift \citepads{2020A&A...642A.150L}, yet slight statistical fluctuations are possible from bin to bin. Therefore, we additionally examine the X-ray intensity of the seven objects in our main dataset with $\Gamma>1.7$ (i.e. those for which we can safely exclude significant X-ray absorption) in narrow redshift bins.

We determined the slope, intercept, and dispersion of the X-ray-to-UV luminosity relation for the subsets of the L20 sample within the redshift range $z=1.825$\,--\,$2.025$ in bins with a step of $\Delta z = 0.05$. We utilised the same fitting method as for the complete L20 sample, namely the Python package \verb|EMCEE|, conducting the regression fit with the normalisation of X-ray and UV luminosities set to the respective median values of the analysed subsets. We then supplemented the subsets of the L20 sample with the sources from our main dataset and repeated the fitting.

\begin{figure*}[b]
   \resizebox{\hsize}{!}
            {\includegraphics[width=\hsize]{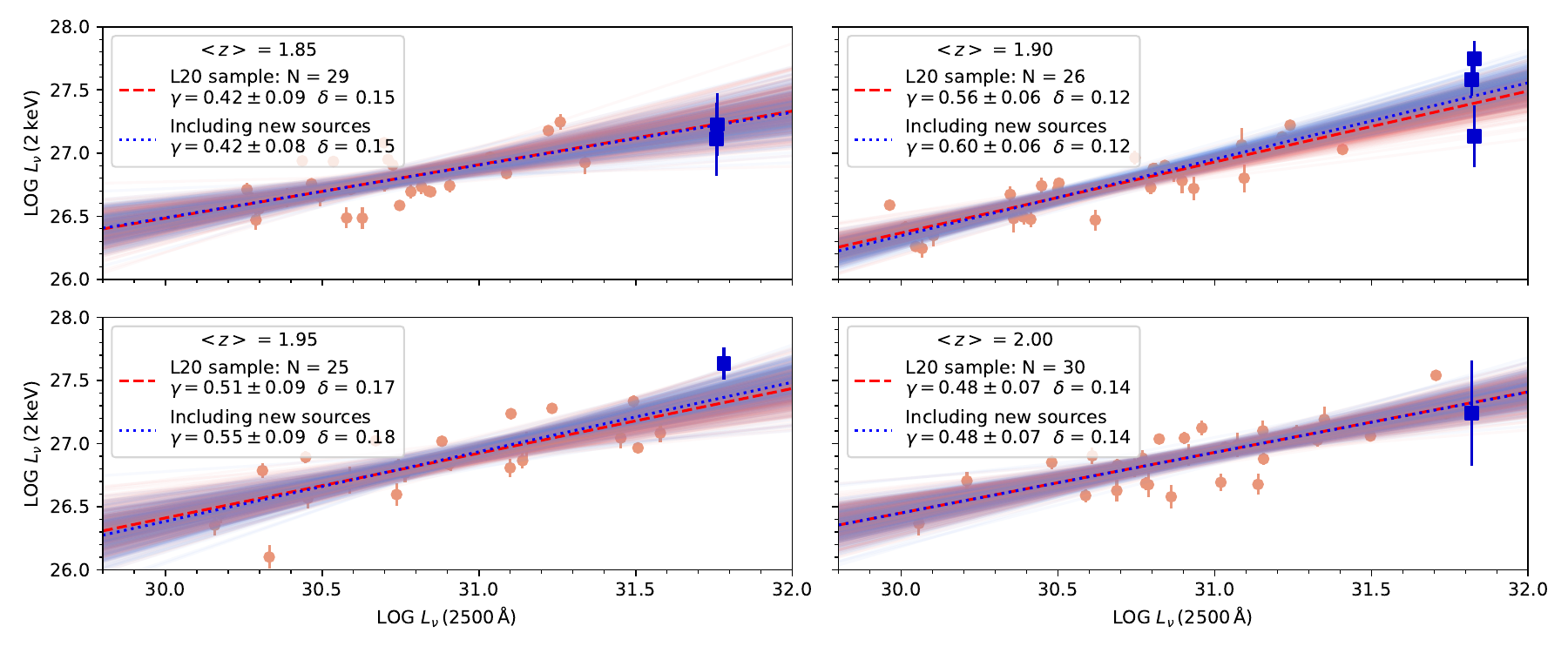}}
      \caption{Rest-frame monochromatic luminosities $L_{\mathrm{X}}$ against $L_{\mathrm{UV}}$ in $\Delta z$=0.05 redshift bins. Light red dots represent quasars from the \citetads{2020A&A...642A.150L} sample, with the relative regression lines shown in dashed red. Blue squares represent seven X-ray steep ($\Gamma>1.7$) quasars in our main dataset, with the updated regression lines shown in dotted blue.}
         \label{fig:relinz}
         \vspace{4cm}
   \end{figure*}

Figure~\ref{fig:relinz} displays the best-fit parameters (slope $\gamma$ with its uncertainty and dispersion $\delta$) of the $L_{\mathrm{X}}$ -- $L_{\mathrm{UV}}$ relation obtained for the aforementioned subsamples. Objects from our main dataset are X-ray `normal' within $\Delta C = 2.706$ and the $2\sigma$ dispersion of the relation in their respective redshift bins. The addition of new sources maintains the same slope of the relation within the error margins for each analysed redshift interval. Meanwhile, the new sources occupy poorly populated areas of the plots, thereby improving the reliability of the fits.

\end{appendix}

\end{document}